\documentclass{article}

\usepackage{algorithm2e}
\usepackage{boxedminipage}

\usepackage{amsmath, amsthm, amssymb, bbm}
\usepackage{graphicx}
\usepackage{verbatim}
\usepackage{natbib}
\usepackage{caption}
\usepackage{subcaption}
\usepackage{fancyvrb}
\usepackage{enumerate}
\usepackage{relsize}
\usepackage{scalerel}
\usepackage{varioref}

 \usepackage{array,multirow,graphicx}
\usepackage{ragged2e}
\usepackage{float}

\usepackage{tikz}
\usetikzlibrary{fit,positioning}

\usepackage{hyperref}
\usepackage[margin=1.5in]{geometry}
\usepackage{chngpage}
\hypersetup{colorlinks,citecolor=blue,urlcolor=blue,linkcolor=blue}

\usepackage{stefan_tex}
\graphicspath{{./figures/}}

\theoremstyle{plain}
\newtheorem{prop}{Proposition}

\newtheorem{theo}[prop]{Theorem}

\theoremstyle{definition}

\theoremstyle{remark}

\date{Draft version \ifcase\month\or
	January\or February\or March\or April\or May\or June\or
	July\or August\or September\or October\or November\or December\fi \ \number%
	\year\ \  }

\author{Xinkun Nie \\ \texttt{xinkun@stanford.edu}
\and Chen Lu \\ \texttt{chenl819@mit.edu}
\and Stefan Wager \\ \texttt{swager@stanford.edu}}

\title{Nonparametric Heterogeneous Treatment Effect Estimation in Repeated Cross Sectional Designs}
\begin{document}

\maketitle
\begin{abstract}
Identifying heterogeneity in a population's response to a health or policy intervention is crucial for evaluating and informing policy decisions. We propose a novel heterogeneous treatment effect estimator in the difference-in-differences design with repeated cross sectional data, where we observe different samples of a population at two time periods separated by the onset of a policy intervention, as well as samples of a population that serves as the control. Our estimator has orthogonality properties that enable fast rates on learning the treatment effect while allowing slower rates for estimating nuisance components. Our proposal shows promising empirical performance across a variety of simulation setups.
\end{abstract}

\section{Introduction}
Difference-in-differences is an increasingly popular observational study design for estimating causal effects from repeated
cross sections \citep[e.g.,][]{angrist2008mostly,bertrand2004much,card1994minimum,lechner2011estimation, obenauer1915effect}. For an overview of applying difference-in-differences method in public health, see  \cite{dimick2014methods}, \cite{gertler2011impact} and \cite{wing2018designing} for a review. Recently, several authors have employed it to study public health policy and implications due to COVID-19 \citep{brodeur2021covid,goodman2020using}.

A standard design with cross sectional data is as follows: we conduct a survey or draw a sample for some outcome (e.g., medical expediture) from two comparable states
(or cities, regions, etc.). The first state later enacts some health policy of interest (e.g., encouragement of Medicare enrollment) and the second state doesn't. We then draw another sample from both states at a time after the health policy has been implemented in the first state. The simplest difference-in-differences estimator assumes global parallel trends: if neither state had enacted the policy, then their trends would have evolved in the same way. It would then attribute any difference in trends between the two states to the effect of the policy change.

In this paper, we focus on flexibly estimating treatment effect heterogeneity in the above design. There are two main challenges. The first challenge involves representing and targeting heterogeneity effectively without assuming specific functional forms. The second challenge involves relaxing the classic assumption of ``global parallel trends," which is unlikely to hold when there is heterogeneity explained by covariates. In particular, states may have changing subgroups of people that exhibit markedly different trends on their own. For example, when studying the effect of enrollment in Medicare on
medical expenditure, we may find that there are subgroups within states (e.g., based on age, income or gender) that have different
baseline trends; then, if the two states under comparison have different proportions of these subgroups, the global parallel trends
assumption immediately becomes questionable. Instead, we may want to control for these covariates, and only assume parallel
trends once we have conditioned on them
\citep[e.g.,][]{ abadie2005semiparametric, acemoglu2001consequences, blundell2004evaluating, heckman1998matching}. 

Throughout the paper, we work in the following formal setup. We observe $n$ independent samples
$(S_i, \, T_i, \, X_i, \, Y_i)$, where the state indicator $S_i \in \cb{0, \, 1}$ denotes whether the $i$-th individual is in the control or exposed state,
$T_i \in \cb{0, \,1}$ denotes the time of the observation (pre- vs. post-intervention), $X_i \in \RR^d$ is a set of potential confounders
and $Y_i \in \RR$ is the outcome of interest. Only samples in the ``exposed'' state and ``post'' time period
get treated, i.e., we can write the treatment or exposure indicator as $W_i = S_iT_i$. Following the potential outcomes framework \citep{imbens2015causal}, let
$Y_i(0)$ and $Y_i(1)$ denote the control and treated potential outcomes and suppose we observe $Y_i = Y_i(W_i)$.
We are interested in estimating the heterogeneous treatment effect  $\tau(x)$, defined as the expected treatment effect
conditional on covariates $X_i = x$ and on $X_i$ treated:
\begin{equation}
	\tau(x) = \EE{Y(1)\cond X_i = x, S_i = 1, T_i = 1} - \EE{Y(0)\cond X_i = x, S_i=1, T_i = 1}.
\end{equation}
Because $\EE{Y(0)\cond X_i=x, S_i = 1, T_i = 1}$ is not observed, we impose a parallel trends assumption conditional on covariates, so that
\begin{equation}
	\label{eq:cond-parallel}
	\begin{split}
		&\EE{Y(0)\cond X_i =x, S_i = 1, T_i = 1} =  \EE{Y(0)\cond X_i =x, S_i = 1, T_i = 0} \\
		&\ \ \ \ \ \ \ \ \ + \EE{Y(0)\cond X_i =x, S_i = 0, T_i = 1} - \EE{Y(0)\cond X_i =x, S_i = 0, T_i = 0},
	\end{split}
\end{equation}
which then allows us to identify the conditional average treatment effect on the treated as follows,
\begin{equation}
	\label{eq:tau_defn}
	\begin{split}
		\tau(x) &= \EE{Y_i \cond X_i = x, \, S_i = 1, \, T_i = 1}
		- \EE{Y_i \cond X_i = x, \, S_i = 1, \, T_i = 0} \\
		&\ \ \ \ \ \ \ \ \ - \EE{Y_i \cond X_i = x, \, S_i = 0, \, T_i = 1}
		+ \EE{Y_i \cond X_i = x, \, S_i = 0, \, T_i = 0}.
	\end{split}
\end{equation}
As discussed in \citet{abadie2005semiparametric}, \eqref{eq:cond-parallel} may be more credible
than the standard parallel trends assumption that holds without conditioning on $X_i$ as it enables
us to control for known sources of confounding.

A classical approach to estimating $\tau(x)$ would be to employ flexible nonparametric modeling of the treatment effect function. A common practice is to employ a two-way fixed effect linear model with interactions for $Y_i$ in terms of $X_i, \, S_i, \, T_i$ of the form
\begin{equation}
	\label{eq:linspec}
	Y \sim \beta_x^\top X + \beta_s^\top X S + \beta_t^\top X T + \beta_{s,t}^\top X ST,
\end{equation} then
interpreting $\hat{\tau}(x) = \hat{\beta}_{s,t}^\top x$ as the treatment effect \citep[e.g.,][]{anzia2011jackie}.
However, the estimator \eqref{eq:linspec} is not
justified by a nonparametric version of the assumption \eqref{eq:cond-parallel} due to the assumed linear functional forms on $\tau(x)$ and on any confounding effects of $X_i$
that affect the outcome $Y_i$ \citep{angrist2008mostly,ding2019bracketing,keele2013much,lechner2011estimation}. These constraining assumptions can be difficult to satisfy in practice.

In this paper, we aim to flexibly estimate the treatment effect function $\tau(X)$ given only the assumption \eqref{eq:cond-parallel}
along with a relevant form of overlap. As a direct consequence of \eqref{eq:cond-parallel}, the data generating process can be written as the following generic
specification: 
\begin{equation}
\label{eq:true_form}
Y_i = b(X_i) + S_i\cdot \xi(X_i) + T_i\cdot \rho(X_i) + T_i\cdot S_i\cdot \tau(X_i) + \varepsilon_i,
\end{equation}
where the joint distribution of $\cb{T_i, \, X_i, \, S_i}$ may be arbitrary and
\begin{equation}
\label{eq:conditional_eff}
\begin{split}
&\xi(x) = \EE{Y_i \cond X_i = x, \, S_i = 1, T_i=0} - \EE{Y_i \cond X_i = x, \, S_i = 0, T_i=0}, \\
&\rho(x) = \EE{Y_i \cond X_i = x, \, S_i = 0,T_i = 1} - \EE{Y_i \cond X_i = x, \, S_i = 0, T_i= 0}, 
\end{split}
\end{equation}
are the conditional effect of $S$ alone and $T$ alone, and $\EE{\varepsilon_i \cond X_i, S_i, T_i} = 0$. We note that all the conditional effect function $b(\cdot), \xi(\cdot), \rho(\cdot), \tau(\cdot)$ can be nonparametric funcitons.

One naive way to estimate $\tau(\cdot)$ in this model is as follows. Recall our expression for $\tau(x)$ in \eqref{eq:cond-parallel}. From that expression, one might attempt to estimate 

\begin{equation}
	g(z) = g(x,s,t) = \EE{Y_i \cond X_i = x, \, S_i = s, \, T_i = t}
\end{equation}
for the four pairs of $s$ and $t$ on the corresponding subsets of the data, and then estimate $\hat{\tau}(x) = \hat{g}(x,1,1) - \hat{g}(x,1,0) - \hat{g}(x,0,1) + \hat{g}(x,0,0)$. However, this approach is often not robust. As an example where this method might fail, consider a high dimensional linear model, $Y_i(s,t) = X_i^\top \beta_{s,t} + \epsilon_i(s,t)$, with $X_i, \beta_{s,t} \in \mathbb{R}^d$, and $\EE{\epsilon_i(s,t) \cond X_i} = 0$. We might consider fitting the Lasso \citep{tibshirani1996regression} for each $\hat{\beta}(s,t)$ separately, and estimate $\hat{\tau}(x) = x^\top (\hat{\beta}(1,1) - \hat{\beta}(1,0) - \hat{\beta}(0,1) + \hat{\beta}(0,0))$. However, the lasso regularizes each $\hat{\beta}(s,t)$ towards $0$ separately, which might result in $\hat{\beta}(1,1) - \hat{\beta}(1,0) - \hat{\beta}(0,1) + \hat{\beta}(0,0)$ being regularized away from $0$, even when $\tau(x) = 0$ everywhere. See \citet{kunzel2019metalearners} and \citet{nie2021quasi} for a similar discussion on the $T$-Learner for the CATE. 

We seek to build a robust estimator for $\tau(\cdot)$. To this end, we start by considering the case where the underlying treatment effect $\tau(x)$ is constant and the only challenge is to eliminate confounding. We start by developing an orthogonal transformation of \eqref{eq:true_form}
that generalizes the transformation of \citet{robinson1988root} for the conditionally linear model.
This representation allows us to build
a transformed regression estimator (TR) to estimate the constant treatment effect. Our TR estimator achieves the parametric $1/\sqrt{n}$
rate of convergence while allowing for slower estimation rates on all nuisance components. The $1/\sqrt{n}$ rate of convergence in the TR estimator allows valid asymptotic confidence interval construction, while we still enable flexible nonparametric estimation on the nuisance components (e.g., without linearity assumptions).  We further discuss the properties of the transformed regression estimator when the underlying linearity assumption is misspecified. We then build upon the transformed regression construction and propose a heterogeneous treatment effect estimator for $\tau(\cdot)$ and show empirically the proposal is advantageous comparing to exisitng baselines.

Compared to the standard difference-in-differences design, we do not assume the panel setting where every individual is observed both before and after the treatment. In that particular case, one could apply any of the existing heterogeneous treatment effect estimators \citep[e.g.,][]{athey2018generalized, hill2011bayesian, kunzel2019metalearners, nie2021quasi} by taking the difference between the outcomes from the two time periods as the new outcome variable. In our setting, we make the less stringent assumption of conditional parallel trends, and allow covariate shifts for the two time periods which makes the setting significantly more challenging. Adapting advances in semiparametric efficiency theory and leveraging flexible nonparametric machine learning methods for this task is our main methodological contribution. 

\section{Related Work}
\label{sec:related-works}
The difference-in-differences approach to treatment effect estimation was popularized by
\citet{card1994minimum}, and has since become ubiquitous in the social sciences.
\citet{angrist2008mostly} and \citet{lechner2011estimation} provide a textbook treatment
and a broad literature review. Building on \citet{abadie2005semiparametric}, we are here most interested
in extensions of classical difference-in-differences methods that leverage covariate information
to make the parallel trends assumption more plausible. Several authors have also recently
extended difference-in-differences analyses in other complementary directions.
\citet{arkhangelsky2018dealing} and \citet{athey2006identification} consider difference-in-differences
type designs where there may be non-additive treatment effects. \citet{abadie2010synthetic}, \citet{arkhangelsky2018synthetic},
\citet{athey2018matrix}, \citet{ben2018augmented},   \citet{xu2017generalized} and \citet{li2019double} develop methods that
can be applied in the panel setting in which the same individuals are observed in both the pre- and post-periods, whereas in our setting we observe separate cross-sectional data in each period.

Most of the existing literature with repeated cross sections, including \citet{abadie2005semiparametric}, \citet{li2019double}, \citet{sant2018doubly}, \citet{chang2020double}, assume that there is no covariate shift across cross-sections from the same
state: they require that the joint distribution of $(X_i, S_i)$ does not vary with $T_i$, i.e., that $(X_i, \, S_i)\indep T_i$.
Our approach does not require this assumption (see Proposition \ref{prop:decomp} and the following comment), as this assumption may be hard to justify with cross-sectional data where we are not able to survey exactly the same people in the pre- and post- periods.
For example, in a ride-sharing application, \citet{lu2018surge} estimates the effects of a dynamic pricing feature on drivers' behaviors
by leveraging a natural experiment where a software bug temporarily disables a dynamic pricing feature for certain drivers.
In this case, we may expect the distribution of the covariates $X_i$ for active drivers varies both with exposure $S_i$ and time $T_i$.  

Methodologically, we build on a large body of work in nonparametric estimation of heterogeneous treatment effects well. One approach is to reduce the ``regularization bias" that might occur. Examples of this line of work include \citet{athey2015machine}, \citet{hahn2017bayesian}, and \citet{shalit2016estimating}. Another approach, the one we choose to adopt, is to develop meta-learning procedures that do not depend on a specific machine learning method. Key examples of such works are \citet{kunzel2019metalearners} and \citet{nie2021quasi}. Our decomposition of $\tau(x)$ is conceptually similar to the orthogonal moments constructions from \citet{robinson1988root}, and more broadly, from \citet{belloni2011square}, \citet{bickel}, \citet{newey1994asymptotic}, \citet{scharfstein1999adjusting}, \citet{van2006targeted} and others. 

In the difference-in-differences design, flexible modeling and estimation that goes beyond the standard two-way fixed effects and linearity assumptions has also drawn considerable interest. \citet{abadie2005semiparametric} considers inverse propensity stratification based methods. Another approach considers more flexible outcome models, see \citep[e.g.,][]{heckman1998matching,meyer1995natural}. Recently, \citet{chang2020double}, \citet{sant2018doubly}
and \citet{zimmert2018efficient} proposed doubly robust variants of the approach of \citet{abadie2005semiparametric}
that also allow for heterogeneity in $\tau(x)$.

\section{An Orthogonal Transformation for the Repeated Cross Sections}
\label{sec:partial_decomp}
The key orthogonality property of our proposed heterogeneous treatment effect estimator relies on a new decomposition for the outcome model \eqref{eq:cond-parallel} motivated by \citet{robinson1988root}. 
The conditional
probabilities of an observation being in state $S_i$ or time period $T_i$ conditionally on $X_i = x$
play a central role in our analysis \citep{rosenbaum1983central}. We write these quantities as
\begin{equation}
\label{eq:propensity}
s(x) = \PP{S_i = 1\cond X_i = x}, \ \ t(x) = \PP{T_i = 1\cond X_i = x}.
\end{equation}
We write $e_{s,t}(X_i) = \PP{S_i = s, T_i = t\cond X_i}$. We also write $m(x) = \EE{Y_i \cond X_i = x}$ for the conditional response function
marginalizing over $T_i$ and $S_i$, and
\begin{equation}
\label{eq:marginal}
\begin{split}
&\varsigma(x) = \EE{Y_i \cond X_i = x, \, S_i = 1} - \EE{Y_i \cond X_i = x, \, S_i = 0}, \\
&\nu(x) = \EE{Y_i \cond X_i = x, \, T_i = 1} - \EE{Y_i \cond X_i = x, \, T_i = 0}, 
\end{split}
\end{equation}
for the conditional effect of $S$ marginalizing over $T$ and $S$ respectively. We write the conditional covariance of $S_i$ and $T_i$ as
\begin{equation}
\label{eq:covariance}
\Delta(x) = e_{1,1}(x) - s(x) t(x).
\end{equation}
Finally, for convenience, we let $Z_i = (X_i, S_i, T_i)$. Given this notation, we can verify the following (the derivation is given in the Supplementary Materials).
\begin{prop}
\label{prop:decomp}
Suppose we have access to an independent and identically distributed sequence of tuples
$Y_i$ and $Z_i = (X_i, \, S_i, \, T_i)$. Under the model \eqref{eq:cond-parallel}, our data-generating distribution admits a representation
\begin{equation}
\label{eq:decomp}
Y_i = m(X_i) + A(Z_i) \nu(X_i) + B(Z_i) \varsigma(X_i) + C(Z_i) \tau(X_i) + \epsilon_i,
\end{equation}
where $\EE{\epsilon_i \cond Z_i} = 0$, and 
\begin{equation}
\label{eq:ABC}
\begin{split}
& A(Z_i) = \p{1 - \frac{\Delta^2(X_i)}{s(X_i)(1 - s(X_i))t(X_i)(1 - t(X_i))}}^{-1}\p{T_i - t(X_i) - \frac{ \Delta(X_i) \p{S_i - s(X_i)}}{s(X_i)(1 - s(X_i))}}, \\
& B(Z_i) = \p{1 - \frac{\Delta^2(X_i)}{s(X_i)(1 - s(X_i))t(X_i)(1 - t(X_i))}}^{-1}\p{S_i - s(X_i) - \frac{ \Delta(X_i) \p{T_i - t(X_i)}}{t(X_i)(1 - t(X_i))}}, \\
&C(Z_i) = S_iT_i -e_{1,1}(X_i) - \p{s(X_i) + \frac{\Delta(X_i)}{t(X_i)}} A(Z_i) -  \p{t(X_i) + \frac{\Delta(X_i)}{s(X_i)}} B(Z_i).
\end{split}
\end{equation}
Furthermore, all terms in the above decomposition are orthogonal in the following sense:
\begin{equation}
\label{eq:orthog_prop}
\begin{split}
&\EE{A(Z) \cond X} = \EE{B(Z) \cond X} = \EE{C(Z) \cond X} = 0, \\
&\EE{A(Z) \cond X, \, S} = \EE{C(Z) \cond X, \, S} = 0, \\
&\EE{B(Z) \cond X, \, T} = \EE{C(Z) \cond X, \, T} = 0,\\
&\EE{B(Z)C(Z) \cond X} = \EE{A(Z)C(Z)\cond X} = 0.
\end{split}
\end{equation}
\end{prop}
The key property of this representation is the orthogonality property \eqref{eq:orthog_prop},
which will enable flexible estimation of treatment effects at parametric rates as discussed in the following section. 
In the setting of \citet{abadie2005semiparametric} and \citet{sant2018doubly}, their assumption gives $T_i \indep S_i \cond X_i$, which implies $\Delta(X_i) = 0$. As an immediate corollary to Proposition \ref{prop:decomp}, this decomposition then simplifies to a functional
form closely reminiscent of Robinson's transformation \citep{robinson1988root}:
\begin{equation}
\label{eq:decomp_simple}
\begin{split}
Y_i &= m(X_i) + (S_i - s(X_i)) \varsigma(X_i) + (T_i - t(X_i)) \nu(X_i) \\
&\ \ \ \ \ \ \ \ \ \ \ \p{S_iT_i - t(X_i)S_i - s(X_i)T_i + s(X_i)t(X_i)} \tau(X_i) + \epsilon_i.
\end{split}
\end{equation}
More generally, we see that when $\Delta(X_i)$ is close to 0, all expressions underlying \eqref{eq:decomp}
and \eqref{eq:ABC} are well-conditioned, and we expect estimation using \eqref{eq:decomp} to be stable.
Conversely, if $T_i$ and $S_i$ are highly correlated conditionally on $X_i$, then
$\Delta^2(X_i) \approx s(X_i)(1 - s(X_i))t(X_i)(1 - t(X_i))$ and \eqref{eq:ABC} could become unstable;
this is as expected, because if $S_i$ and $T_i$ are highly correlated, then we do not expect their
interaction effect to be well identified.

Finally, we note that all nuisance components in the decomposition above are marginal quantities,
and thus can be estimated using all of the data. This property is desirable for empirical performance as it is
more data efficient when we need to estimate them in a small-sample regime.

\section{The Transformed Regression Estimator}
\label{sec:const-eff}
As a building block of our proposed heterogeneity treatment effect estimator, we first consider estimation in a setting where
the treatment effect itself is constant $\tau(x) = \tau$ in the representation \eqref{eq:decomp}, but all other
nuisance components defined above, i.e., $m(x)$, $\nu(x)$, $\varsigma(x)$, $s(x)$, $t(x)$ and $\Delta(x)$,
may vary with $x$. The standard approach to estimating $\tau$ in this setting is to write a two-way fixed effect model of the form
\begin{align}
	\label{eq:fixed-effect}
	Y\sim \beta_x^\top X + \beta_s S+\beta_t T + \beta_{s,t} ST
\end{align}
and to interpret the coefficient on $ST$ as an estimate of the treatment effect. However, as shown in our experiments,
this simple linear regression-based approach to treatment effect estimation may be severely biased in the setting where
the linear model \eqref{eq:fixed-effect} is misspecified.

Here, we propose the \textbf{transformed regression (TR)} estimator with cross-fitting (shown in Algorithm \ref{alg:TR}). The method is based on the decomposition \eqref{eq:decomp}, which is motivated by a decomposition used by \citet{robinson1988root} to estimate parametric components in partial linear models. Robinson's decomposition has also been used in many other recent works, such as in \citet{athey2018generalized} for causal forests, \citet{robins2004optimal} for G-estimation, as well as in \citet{chernozhukov2016double} and \citet{zhao2017selective}. The transformed regression estimator, motivated by Robinson, also has good theoretical properties. In Theorem \ref{theo:const_tau}, we show that the transformed regression estimator is $\sqrt{n}$-consistent and asymptotically normal under considerably more generality than simply running an OLS regression with the model \eqref{eq:fixed-effect}. We note that cross-fitting helps avoid overfitting of the estimates and also serves as a proof technique to show $\sqrt{n}$-consistency on $\tau$. 
Having $\sqrt{n}-$consistency on $\tau$ enables us to build valid asymptotic confidence intervals for $\tau$, while we still allow nonparametric estimation on the nuisance components without imposing linearity assumptions such as in \eqref{eq:fixed-effect}.
The proof of the theorem is in the Supplementary Materials.

\RestyleAlgo{boxruled}
\LinesNumbered
\begin{algorithm}[!htbp]
  \caption{\textbf{Transformed Regression Estimator (TR)}\label{alg:TR}}
  Split the data into $Q$ roughly equal folds, $\mathcal{I}_1$, $\mathcal{I}_2$, ..., $\mathcal{I}_K$, with $K$ fixed, to be used for cross-fitting. 

For each fold ${{\mathcal{I}_{k}}}$, fit $\hat{m}^{-{\mathcal{I}_{k}}}(x), \hat{s}^{-{\mathcal{I}_{k}}}(x), \hat{t}^{-{\mathcal{I}_{k}}}(x)$ and $\hat{e}^{-{\mathcal{I}_{k}}}_{1,1}(x)$, with data not in ${{\mathcal{I}_{k}}}$, using any supervised learning method for prediction accuracy (the superscript of $-{\mathcal{I}_{k}}$ denotes using data not in the $k$-th fold).

Estimate $\hat{\nu}^{-{\mathcal{I}_{k}}}(x)$ as a heterogeneous ``treatment effect" of $S_i$ while ignoring $T_i$; and estimate $\hat{\varsigma}^{-{\mathcal{I}_{k}}}(x)$ as a ``treatment effect" of $T_i$ ignoring $S_i$. Both can leverage methods designed for heterogeneous treatment estimation in the single cross-section case.

Construct $\hat{A}^{-{\mathcal{I}_{k}}}(z)$, $\hat{B}^{-{\mathcal{I}_{k}}}(z)$, $\hat{C}^{-{\mathcal{I}_{k}}}(z)$ and $\hat{\Delta}^{-{\mathcal{I}_{k}}}(x)$, where $z = (x,s,t)$, using the estimated nuisance parameters following \eqref{eq:ABC}. Then, for $j\in {{\mathcal{I}_{k}}}$, obtain point estimates $\hat{C}^{-{\mathcal{I}_{k}}}(Z_j)$ and
\begin{equation}\label{eq:H}
\hat{H}^{-{\mathcal{I}_{k}}}(Z_j) = Y_j - \p{\hat{m}^{-{\mathcal{I}_{k}}}(X_j) + \hat{A}^{-{\mathcal{I}_{k}}}(Z_j) \hat{\nu}^{-{\mathcal{I}_{k}}}(X_j) + \hat{B}^{-{\mathcal{I}_{k}}}(Z_j) \hat{\varsigma}^{-{\mathcal{I}_{k}}}(X_j)}.
\end{equation}

Run OLS on $\hat{H}^{{-\mathcal{I}_{k}}}(Z_j)$ against $\hat{C}^{-{\mathcal{I}_{k}}}(Z_j)$ to produce
\begin{equation}\label{eq:ols_ith}
\hat{\tau}^{-\scaleto{\mathcal{I}_k}{5pt}} = \frac{\sum_{j\in \mathcal{I}_k} \hat{H}^{-\mathcal{I}_{k}}(Z_j) \hat{C}^{-\mathcal{I}_{k}}(Z_j)}{\sum_{j\in \mathcal{I}_k} \hat{C}^{-\mathcal{I}_{k}}(Z_j)^2}
\end{equation}

Combine predictions from different folds ${{\mathcal{I}_{k}}}$:
\begin{equation}\label{eq:TR}
\hat{\tau}_{TR} = \sum_{i=k}^{K} \frac{|\mathcal{I}_k|}{n} \hat{\tau}^{-\scaleto{\mathcal{I}_k}{5pt}} 
\end{equation}
\end{algorithm}

\begin{theo}
\label{theo:const_tau}
Under the conditions of Proposition \ref{prop:decomp}, suppose furthermore that $\tau(x) = \tau$ is
constant and that the following conditions hold:
\begin{enumerate}
\item Overlap: the conditional probabilities $e_{s,t}(x)$ are bounded away from $0$
by some small $\eta > 0$ for all values of $t$, $s$ and $x$.
\item Consistency: for any estimated nuisance parameter $\hat{\mu}(x)$, such as $\hat{m}(x)$, $\hat{\nu}(x)$ and $\hat{\varsigma}(x)$, we have that:
$$
\sup_{x}|\hat{\mu}(x) - \mu(x)| \rightarrow_p 0
$$
\item Risk decay: for any estimated nuisance parameter $\hat{\mu}(x)$, we have:
$$
\mathbb{E}\left[(\hat{\mu}(x) - \mu(x))^2\right] = o\left(\frac{1}{\sqrt{n}}\right)
$$
\item Boundedness: all the nuisance parameters are uniformly bounded:
$$
\sup_x |\mu(x)| \textrm{ , } \enspace \sup_x|\hat{\mu}(x)| < M
$$
for some constant $M < \infty$.
\end{enumerate}
Then, writing $\hat{\tau}_{TR}$ as the transformed regression estimater obtained using Algorithm \ref{alg:TR}, and $\hat{\tau}^*$ as the transformed regression estimator with oracle nuisance parameters, we have
$$
\sqrt{n}\p{\hat{\tau}_{TR} - \hat{\tau}^*} \xrightarrow{p} 0, \ \ \sqrt{n}\p{\hat{\tau}^* - \tau} \xrightarrow[]{d} \mathcal{N}\p{0,V_{TR}},
$$
where 
\begin{equation}
V_{TR} = \frac{\EE{\sigma^2(z) C^2(z)}}{\EE{C^2(z)}^2}.
\end{equation}
and $\sigma(z)^2 = \Var{\epsilon_i\cond Z_i = z}$.
In the case when $T_i \indep S_i \cond X_i$, and $\sigma^2(z) = \sigma^2$ is constant, the expression for $V_{TR}$ simplifies to
$V_{TR} = \sigma^2/\EE{t(x)(1-s(x))t(x)(1-s(x))}$.
\end{theo}
In step 3 of Algorithm \ref{alg:TR}, $\varsigma(x)$ and  $\nu(x)$ can be estimated with methods for heterogeneous treatment effect estimation.  In our simulations, we use causal forests \citep{athey2018generalized}, but we note that other estimators in the literature can also be used here
 \citep[e.g.,][]{athey2018generalized, hill2011bayesian, kunzel2019metalearners, nie2021quasi}.

If we ever want to use the transformed regression estimator which assumes a constant treatment effect, it is important to understand how it
behaves under misspecification. Interestingly, as shown in Proposition \ref{prop:weight}, even when $\tau(x)$ is not constant, the transformed regression estimator converges to a weighted average of $\tau(x)$ with positive weights, mirroring the findings in \citet{crump2009dealing}
and  \citet{li2018balancing}. The proof for Proposition \ref{prop:weight} is found in the Supplementary Materials.

\begin{prop}
	\label{prop:weight}
	If we use the transformed regression estimator from Algorithm \ref{alg:TR}, and the conditions from Theorem \ref{theo:const_tau} are satisfied, then
	\begin{equation}\label{eq:weighted_tau}
	\sqrt{n}(\hat{\tau}_{TR} - \bar{\tau}) \xrightarrow{d} \mathcal{N}(0,V_{TR}),\ \
	\bar{\tau} = \frac{\EE{C^2(z)\tau(x)}}{\EE{C^2(z)}},
	\end{equation}
where  $V_{TR}$ is the same variance term from Theorem \ref{theo:const_tau}.
\end{prop}

We also note that when in the setting of \citet{abadie2005semiparametric},
where $T_i \indep S_i \cond X_i$ and so $\Delta(X_i) = 0$, the above simplifies to:
$$
\bar{\tau} = \frac{\EE{[s(x)(1-s(x))t(x)(1-t(x))]\cdot\tau(x)}}{\EE{s(x)(1-s(x))t(x)(1-t(x))}}
$$
When $\tau(x)$ is not constant, the transformed regression estimator can thus be thought of as a weighted mean of the treatment effect, where more weight is given to the data points that are likely to appear with all four $(S_i,T_i)$ pairs.

\section{Estimating Treatment Heterogeneity with Repeated Cross Sectional Data}
\label{sec:hte}
In this section, we relax the assumption from the previous section that the underlying treatment effect is constant, and aim to estimate the heterogeneous treatment effect. We propose a flexible nonparametric estimator in the difference-in-differences setup that draws inspiration from recent advances in heterogeneous treatment effect estimation in the single cross-section case. 

We adapt our estimator of a constant causal parameter $\tau$
into an estimator for a heterogeneous treatment function $\tau(x)$, by turning the estimation equation underlying
the former estimator into a loss function. The R-learner \citep{nie2021quasi} follows this strategy to derive a heterogeneous treatment effect estimator from Robinson's decomposition \citep{robinson1988root} in the setting of a single cross section. We follow the same strategy here with repeated cross sectional data. Using the decomposition from
\eqref{eq:decomp}, we can estimate treatment effect heterogeneity with the following algorithm \textbf{R-DiD}:

\RestyleAlgo{boxruled}
\LinesNumbered
\begin{algorithm}[t]
	\caption{\textbf{Heterogeneous Treatment Effect Estimation with Cross Sectional Data (R-DiD)}\label{alg:h_did}}
	Split the data into $K$ roughly-equal folds $\mathcal{I}_1$, ..., $\mathcal{I}_K$ for cross-fitting.
	
	Following steps 2 to 4 of Algorithm \ref{alg:TR}, estimate the nuisance parameters $\hat{m}^{-{\mathcal{I}_{k}}}(x)$, $\hat{\varsigma}^{-{\mathcal{I}_{k}}}(x)$, $\hat{\nu}^{-{\mathcal{I}_{k}}}(x))$, $\hat{A}^{-{\mathcal{I}_{k}}}(z)$, $\hat{B}^{-{\mathcal{I}_{k}}}(z)$, $\hat{C}^{-{\mathcal{I}_{k}}}(z)$ using data not in the $k$-th fold. Also following step 4 of Algorithm \ref{alg:TR}, produce point estimates $\hat{C}^{-{\mathcal{I}_{k}}}(Z_j)$ and $\hat{H}^{-{\mathcal{I}_{k}}}(Z_j)$ according to \eqref{eq:H} for $j\in \mathcal{I}_k$.
	
	Estimate the cross-fitted heterogeneous treatment effect $\hat{\tau}^{(-\scaleto{\mathcal{I}_k}{5pt})}(\cdot)$ for data in $\mathcal{I}_k$ as
	\begin{equation}
	\label{eq:non_const_regression}
	\begin{split}
	\hat{\tau}^{(-\scaleto{\mathcal{I}_k}{5pt})}(\cdot) &= \argmin_\tau \bigg\{ \frac{1}{|\mathcal{I}_k|} \sum_{j\in \mathcal{I}_k} \left( \hat{H}^{-{\mathcal{I}_{k}}}(Z_j) - \hat{C}^{-{\mathcal{I}_{k}}}(Z_j) \tau(X_j) \right)^2 + \Lambda_n(\tau(\cdot)) \bigg\},
	\end{split}
	\end{equation}
	where $\Lambda_n(\cdot)$ is some regularization term. For $j\in\mathcal{I}_k$, use $\hat{\tau}^{(-\scaleto{\mathcal{I}_k}{5pt})}(X_j)$ as the estimate for $\tau(X_j)$.
\end{algorithm}

Recall from the last section that when the treatment effect $\tau(x)$ is contant, the transformed regression estimator achieves $\sqrt{n}$ rate for estimating the treatment effect parameter $\tau$. When $\tau(x)$ is a nonparametric function, we can no longer achieve $\sqrt{n}$ parametric rates. Instead, we aim to show a 		quasi-oracle result, i.e., even if estimating the nuisance components $m(x)$, $\nu(x)$, $\varsigma(x)$, $\Delta(x)$, $s(x)$ and $t(x)$ have a slow convergence rate, we can still achieve fast nonparametric rates on the treatment effect function $\tau(x)$ as if we had known these nuisnace components perfectly. See \citet{chernozhukov2017double, luedtke2016super, van2003cross} for similar developments. In this work, we leverage the recent result from \citet{foster2019orthogonal} that generalizes the quasi-oracle bounds in \citet{nie2021quasi}. 

In particular, for any treatment effect function $\tilde{\tau}$ and nuisance function $\tilde{\mu}$ where the nuisance function $\mu(x)$ includes $m(x)$, $\nu(x)$, $\varsigma(x)$, $\Delta(x)$, $s(x)$ and $t(x)$, define the loss function 
\begin{equation}
	L(\tilde{\tau}, \tilde{\mu}) = \EE{(Y - \tilde{m}(X) - \tilde{A}(Z) \tilde{\nu}(X) - \tilde{B}(Z) \tilde{\varsigma}{X} - \tilde{C}(Z)\tilde{\tau})^2 }.
\end{equation}
Suppose in Algorithm 2, instead of leveraging cross-fitting to learn nuisance components, we use sample splitting, i.e. we split the data in half, and use the first half to learn all the nuisance components $\hat{\mu}$, and use the second half to estimate $\hat{\tau}$. The following then holds as a direct consequence of Theorem 1 in \citet{foster2019orthogonal}, 
\begin{theo}
	\label{theo:hte-rate}
	Suppose the conditional probabilities $e_{s,t}(x)$ are bounded away from 0 by some constant $\eta>0$ for all values of $t,s$ and $x$, and the outcome $\abs{Y}$ is bounded by $M$. Suppose further that $K=2$ in Algorithm \ref{alg:h_did}, and let $\hat{\tau} = \hat{\tau}^{(-2)}$ be the treatment effect estimate on the first fold. Suppose there exist rate functions $r_n$ such that 
	\begin{equation}
   \Norm{\hat{\mu} - \mu}_{L_2}^4 = O_p(r(n))
	\end{equation} and
	\begin{equation}
    L(\hat{\tau},\hat{\mu}) - L(\tau, \hat{\mu})= O_p(r(n)), \label{eq:lossrate}
\end{equation}
then 
\begin{equation}
	\Norm{\hat{\tau} - \tau}_{L_2(P)}^2 = O_p(\eta^{-16} M^4 r(n))
\end{equation}
for all nuisance parameter $\mu(x)$ including $m(x)$, $\nu(x)$, $\varsigma(x)$, $\Delta(x)$, $s(x)$ and $t(x)$.
\end{theo}
The theorem above shows that the required learning rate on the nuisance components is only at a 4th-order growth rate compared to the learning rate on the target parameter $\tau$. Note that if the nuisance components $\mu$ were known,  \eqref{eq:lossrate} would be in terms of $L(\hat{\tau},\mu) - L(\tau, \mu)$. We refer readers to Section 4 in \citet{foster2019orthogonal} to see that in the case of empirical risk minimization, the cost to relate these quantities gets absorbed, and the same conclusion holds. The proof of the theorem relies on the orthogonality results from Proposition \ref{prop:decomp} and is included in the Supplementary Materials.

\section{Simulation Study}

In this section, we test the validity of our methods in a variety of simulation setups where the true underlying treatment effect can be both constant and non-constant. In the simulations, we generate $n$ i.i.d. samples $X_i$ of dimension  $p$ from some underlyng distribution $P$; the pre/post-treatment time indicator $T_i$ and the state indicator $S_i$ are generated with a multinomial distribution over the four pairs $(T_i, S_i) \in \{(1,1), (1, 0), (0,1), (0,0)\}$. The outcomes $Y_i$ are generated from \eqref{eq:true_form} with $\epsilon_i \cond X_i \sim \mathcal{N}(0,1)$.
Next, we present results in the case where the underlying treatment effect is constant as well in the case where it is heterogeneous. 

We compare five methods in estimating heterogeneous treatment effects, four of which serve as baselines. \textbf{OLS} is as shown in \eqref{eq:linspec}. The \textbf{T}-Learner runs a separate regression for each of the four conditional quantities in \eqref{eq:cond-parallel} with a regression forest, and follows  \eqref{eq:cond-parallel} to build the estimate $\hat{\tau}$ from the four separate regressions. The \textbf{CF-time} learner runs a causal forest to learn the time-wise treatment effect in the treated location, i.e. $$\EE{Y_i\cond X_i=x, S_i=1, T_i=1} - \EE{Y_i\cond X_i=x, S_i=1, T_i=0},$$ and then runs another causal forest to learn the time-wise treatment effect in the control location, i.e. $$\EE{Y_i\cond X_i=x, S_i=0, T_i=1} - \EE{Y_i\cond X_i=x, S_i=0, T_i=0},$$ and subtracts the two estimates. On the other hand,  The \textbf{CF-state} learner runs a causal forest to learn the state-wise treatment effect in the treated time period, i.e. $$\EE{Y_i\cond X_i=x, S_i=1, T_i=1} - \EE{Y_i\cond X_i=x, S_i=0, T_i=1} ,$$ and then runs another causal forest to learn the time-wise treatment effect in the control period, i.e. $$\EE{Y_i\cond X_i=x, S_i=1, T_i=0} - \EE{Y_i\cond X_i=x, S_i=0, T_i=0},$$ and subtracts the two estimates. 

We compare the above four baselines with the \textbf{R-DiD} estimator outlined in Algorithm \ref{alg:h_did}. In particular, we note that causal forests \citep{athey2018generalized} can be understood as an instantiation of the R-learner with random forests, and can be used in Step 3 of Algorithm \ref{alg:h_did}.\footnote{For more discussions on the connection between the causal forests and the R-learner, see Section 1.3 in  \citep{athey2019estimating}.}
All regressions in all of the methods under comparison are implemented with regression forests from the package \texttt{grf} \citep{athey2018generalized}. In Algorithm \ref{alg:h_did}, we take $\hat{H}$ as the ``outcome" and $\hat{C}$ as the ``treatment" in a causal forest. Along with \textbf{CF-time} and \textbf{CF-state}, they are all implemented with the package \texttt{grf}.
We consider the following four setups:
\bigbreak
\noindent
\textbf{Setup A} $X_i\sim \mathcal{N}(0,\textrm{I}_{d\times d})$; easy treatment effect $\tau(x) =x_4+  0.5 x_5$; easy conditional effects $\rho(x) = 1/(1+\exp(x_3)) + 4x_5^2$, $\xi(X) = 1/(1+\exp(x_4)) + 3x_6^2$; easy baseline $b(x) = \max(x_1 + x_2, 0) + 4x_6^2$; constant propensity for $S_i$: $s(x) =  0.6$, but constant propensity for $T_i$, $t(x) = 0.4$, and $S_i$ is independent from $T_i$.

\bigbreak
\noindent
\textbf{Setup B} $X_i\sim \mathcal{N}(0,\textrm{I}_{d\times d})$; easy treatment effect $\tau(x) = 0.5(x_1+x_2+x_3)$; challenging conditional effects $\rho(x)=5(sin(x_1x_2\pi) + 2(x_3-0.5)^2)$ and $\xi(x)=5(sin(x_1 x_2\pi ) + 2x_5^2)$. There is no baseline effect, i.e. $b(x)= 0$; propensities are constant $s(x) = t(x) = 0.5$, with $S_i$ and $T_i$ independent.
\bigbreak

\noindent
\textbf{Setup C} $X_i\sim \mathcal{N}(0,\textrm{I}_{d\times d})$; there exists no time or state effect: $\rho(x) = \xi(x) = 0$, but highly correlated baseline effect $b(x) = 2 \sin(1.5x_1)$ and propensities, where $e_{1,1}(x) = 0.5 + 0.5(1-6\eta)\sin(1.5x_1)$, and $e_{s,t}(x) = (1-e_{1,1}(x))/3$, for $(s,t) \not= (1,1)$; $\tau = 1$.

\bigbreak

\noindent
\textbf{Setup D} $X_i\sim \mathcal{N}(0,\textrm{I}_{d\times d})$; easy non-constant treatment effect $\tau(x) = 3x_1 + 2x_4$; easy conditional effects $\rho(X) = 2x_5$, $\xi(X) = 0$; easy baseline $b(x) = \max(x_1 + x_2 + x_4+x_6, 0)$; non-constant propensity for $S_i$: $s(x) = \min(\max(\eta, (1 / (1+\exp(-0.5 x_3)))), 1-\eta)$, and non-constant propensity for $T_i$, $t(x) = \min(\max(\eta, (1 / (1+\exp(-0.5 x_2)))), 1-\eta)$, and $S_i$ is independent from $T_i$. This is a well-specified setup for OLS.
\bigbreak

\begin{table}[t]
	\centering
	\begin{tabular}{cccccccc}
		\hline
		setup & n & p & R-DiD & CF-time & CF-state & T & OLS \\ 
		\hline
		A & 1000 &   6 & \bf 1.01 & 3.83 & 3.29 & 7.8 & 15.05 \\ 
		A & 1000 &  12 & \bf 1.11 & 4.7 & 3.56 & 8.64 & 23.85 \\ 
		A & 2000 &   6 & \bf 0.61 & 2.53 & 1.93 & 5.09 & 7.75 \\ 
		A & 2000 &  12 & \bf 0.67 & 2.7 & 2.02 & 5.63 & 11.61 \\ 
		B & 1000 &   6 & \bf 2.69 & 14.1 & 15.18 & 40.8 & 38.66 \\ 
		B & 1000 &  12 & \bf 2.78 & 15.25 & 19.55 & 49.38 & 64.63 \\ 
		B & 2000 &   6 & \bf 2.07 & 8.8 & 8.71 & 32.56 & 18.19 \\ 
		B & 2000 &  12 & \bf 1.66 & 10.22 & 9.73 & 37.25 & 28.92 \\ 
		C & 1000 &   6 & \bf 0.03 & 0.04 & 0.04 & 0.27 & 0.55 \\ 
		C & 1000 &  12 & \bf 0.02 & 0.03 & 0.04 & 0.29 & 0.78 \\ 
		C & 2000 &   6 & \bf 0.02 & \bf 0.02 & \bf 0.02 & 0.22 & 0.36 \\ 
		C & 2000 &  12 & \bf 0.02 & \bf 0.02 & \bf 0.02 & 0.21 & 0.46 \\ 
		D & 1000 &   6 & 1.92 & 1.64 & 2.73 & 2.34 & \bf 0.18 \\ 
		D & 1000 &  12 & 2.37 & 1.97 & 3.11 & 2.78 & \bf 0.32 \\ 
		D & 2000 &   6 & 1.17 & 1.01 & 1.94 & 1.65 & \bf 0.09 \\ 
		D & 2000 &  12 & \bf 1.34 & 1.45 & 1.35 & \bf 1.34 & 1.4 \\ 
		\hline
	\end{tabular}
	\caption{Simulation results on mean squared error on an independent test set comparing our proposal \textbf{R-DiD} against four other baselines in different simulation setups, with varying training sample size $n$ and dimensions $p$. Results are averaged across 200 independent runs.}
	\label{tab:hte-sim}
\end{table}
The results of the simulations are shown in Table \ref{tab:hte-sim}. For each of the $n$ values, we draw $n$ training data points, and another separate $n$ testing data points. We report results on this indepedent testing set. We run the experiments 200 times, and the mean squared error of each algorithm is shown below. Our proposal {\bf R-DiD} performs well, while in simulation setup D, we see that OLS performs particularly well given it's a well-specified setup. However, in practice it is less conceivable to have a well-specified setup, and our algorithm shines due to its flexibility to model nonparametrically and robustness towards estimation errors in nuisance components. 

\section{Application}
\label{sec:real-app}
To test our methods in practice, we revisit a study from \citet{angrist2008rural} on the effect of import restrictions on self-employment incomes. The context is as follows: Columbia was one of the major suppliers of cocaine to North America and Europe before the $2000$s. Before $1994$, Columbia relied on coca leaf supplies from Bolivia and Peru, which it then refined to produce cocaine. Starting from $1994$, a series interdictions made by the United States and local militaries disrupted the air-bridge that brought the coca leaves to Columbian refiners. As a result, coca cultivation shifted to Columbia's rural areas. The study from \citet{angrist2008rural} then examines, among other things, the effect of the restriction of coca import on self-employment incomes in Columbia. The authors define self-employment income as income from individual short-term contract, from the sale of domestically produced goods, and from agricultural productions. The authors conclude that the decrease in import has a positive effect on the self-employment incomes.

The dataset includes repeated cross-section survey data and contains the following information about individuals: gender, age, number of family members, immigrant status, marital status, and whether they lived in rural or urban areas, which we use as covariates $X_i$. The individuals come from one of three coca-growing regions  (Bolivar, Cauca and Narino), or one of thirteen non-growing regions (Atlantico, Sucre, Cordoba, Santander, Boyaca, Caldas, Risaralda, Quindio, Tolima, Huilda, Antioquia, Choco, Valle de Cauca); see map in \citet{angrist2008rural} for details. The dataset also includes a number of demilitarized zones, which we omit in our
analysis.\footnote{It was suggested that being a demilitarized zone may have an effect on the incomes of people from that region. Moreover, the demilitarized zones were exclusively growing regions. If we were to include the indicator of whether an individual is from a demilitarized region as one of our covariates, all such individuals will be in the treated group, hence violating overlap.} Individuals from growing regions are classified as exposed, with $S_i=1$, and those in non-growing regions classified as non-exposed, with $S_i=0$, because the increase in coca production could only benefit those in growing regions. As for time periods, we take $1993$ as the pre-treatment period, with $T_i = 0$; because air interdictions occured throughtout $1994$, we take $1995$ as the post-treatment period, with $T_i = 1$. The outcomes are the log self-employment incomes, $Y_i$. 
\begin{figure}[t]
	\centering
	\includegraphics[width=0.6\textwidth]{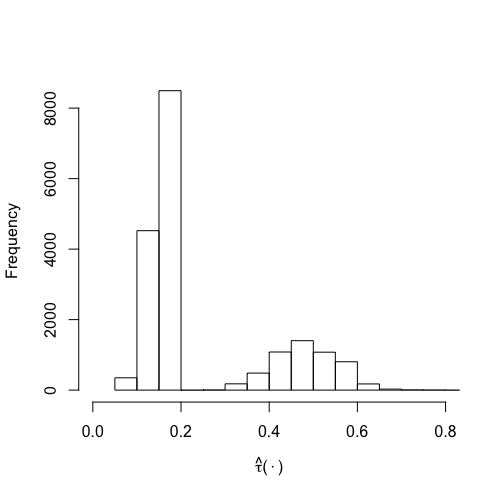}
	\caption{Histogram of $\hat{\tau}(\cdot)$ fitted using Algorithm \ref{alg:h_did}  on the dataset from \citet{angrist2008rural}. We see that there appears to be heterogeneity in the treatment: the treatment effects lie in two groups, one around $0.1$ and the other around $0.5$. Our results in Table \ref{tab:real_sims_urban} suggest that these two groups are treatment effects for people living in urban and rural areas respectively.}
	\label{fig:real_tau_hat}
\end{figure}

We first fit the treatment effect function $\tau(x)$ using the heterogeneous effect estimator $\textbf{R-DiD}$ as in Algorithm \ref{alg:h_did}. Figure \ref{fig:real_tau_hat} provides a histogram of the estimated treatment effects. We see that there appears to be heterogeneity in the treatment, as the histogram exhibits two distinct masses.

Beyond heterogeneity estimates, we compare a few different methods for the estimation of average treatment effects in this setup. In Appendix \ref{app: aipw}, we describe the augmented inverse propensity estimator (\textbf{AIPW}) that builds upon the heterogeneous treatment effect estimates from Algorithm \ref{alg:h_did}. An alternative approach is \textbf{AMLE}, which builds upon the Augmented Minimax Linear Estimation (AMLE) approach of \citet{hirshberg2018balancing} by constructing balancing weights. See an earlier working paper \citep{lu2019robust} for details on this estimator and comparison with \textbf{AIPW}. We also compare our \textbf{TR} estimator that assumes constant treatment effects, as well as \textbf{Sample Means} by taking the empirical means of \eqref{eq:tau_defn}, and the standard \textbf{OLS} as in \eqref{eq:fixed-effect}.\footnote{Note that OLS is the method used in \cite{angrist2008rural}, except we use the data in 1995 as the post-treatment period, but \cite{angrist2008rural} use the data from 1995 up until 2000 as the post-treatment period and assumed a fixed effect for each of the year-treatment interaction.} See Table \ref{tab:real_sims} for result comparisons. 
\tabcolsep = 0.1cm
\begin{table}[t]
	\begin{center}
		\begin{tabular}{|l|cc|}
			\hline
			estimator  & estimate  & std. err  \\ 
			\hline
			Sample Means &0.279 &0.059 \\
			OLS &0.274 &0.031 \\
			TR  &0.275  & 0.031  \\
			AMLE &0.234  & 0.031  \\
			AIPW &0.258  & 0.031  \\
			\hline
		\end{tabular}
		\caption{Average treatment effect estimates in the difference-in-differences setup with data from \citet{angrist2008rural}, where $1993$ is the pre-treatment year and $1995$ is the post-treatment year. The first three methods assume a constant treatment effect $\tau(x) = \tau$, and the latter two allow for treatment heterogeneity. The OLS method is most similar to the method used in \citet{angrist2008rural}. We see that AIPW and AMLE, which allow for heterogeneity in the treatment effects, both obtain a lower point estimate than the other methods, which all assume the underlying effect is constant. This may suggest that there exists heterogeneity in the treatment effects.}
		\label{tab:real_sims}
	\end{center}
\end{table}

Table \ref{tab:real_sims} shows the results for estimating average treatment effects. There exists a non-trivial difference between the estimated effect from methods that allow for heterogeneity in $\tau(x)$ and those that do not, which suggests that there is a weighting effect going on when the likelihood of state and treatment time indicators vary with covariates. All the methods suggest that there is a positive effect of the air interdictions on the log self-employment income, as found in \citet{angrist2008rural}. Recall that, when the treatment effect is non-constant, {\bf TR} obtains estimates of weighted average treatment effects, given by Proposition \ref{prop:weight}. Because {\bf TR} is giving significantly different results from the non-constant effect methods, we suspect that in this dataset the treatment effect varies with covariates. 
We also note that, in this case, the {\bf OLS} estimator is fairly closely aligned with the {\bf TR} estimator, suggesting that assuming linear nuisance
components did not have too big an effect on our estimation of $\tau$. However, it may have been difficult to argue a-priori that
the linear specification used by {\bf OLS} would be innocuous here.

As further evidence of treatment heterogeneity, Table \ref{tab:real_sims_urban} shows average
effects obtained by {\bf AMLE}, separately for urban and rural regions. Thus, estimates provided by the transformed regression estimator should not necessarily be interpreted as average treatment effects, and may instead better be interpreted as targeting a weighted estimand following the discussion in Section \ref{sec:const-eff}.

\tabcolsep = 0.1cm
\begin{table}[t]
\begin{center}
\begin{tabular}{|l|cc|}
\hline
 estimand  & estimate  & std. err  \\ 
\hline
 ATE Urban &0.102  & 0.035  \\
 ATE Rural  &0.582  &0.057  \\
\hline
\end{tabular}
\caption{This table gives evidence that suggests the existence of heterogeneity in the treatment effect from the dataset of \citet{angrist2008rural}. The first row is obtained by running the AMLE method on only individuals who live in the urban area; the second row comes from running AMLE on individuals who live in the rural area. As shown, there seems to be a much larger treatment effect on individuals living in rural areas than on individuals in urban areas.} 
\label{tab:real_sims_urban}
\end{center}
\end{table}

As a final sanity check, we also run a placebo analysis with years $1998$ and $2000$ as pre- and post- treatment period, and we use our methods to check if there was an effect in these years. It is encouraging that all the methods suggest there is a negligible effect on the treatment effect, as shown in Table \ref{tab:real_sims_control}.
\tabcolsep = 0.1cm
\begin{table}[!htbp]
	\begin{center}
		\begin{tabular}{|l|cc|}
			\hline
			estimator  & estimate  & std. err  \\ 
			\hline
			Sample Means &-0.009 &0.047 \\
			OLS &0.004 &0.023 \\
			TR &0.012  & 0.023  \\
			AMLE&0.008  & 0.023  \\
			AIPW &0.007  & 0.023  \\
			\hline
		\end{tabular}
		\caption{Treatment effect estimates with data from \citet{angrist2008rural}, where $1998$ is the pretreatment year and $2000$ is the post treatment year. All the methods suggest that there is negligible treatment effect.}
		\label{tab:real_sims_control}
	\end{center}
\end{table}

\pagebreak

\appendix 
\begin{center}
	\textbf{\LARGE Appendix}
\end{center}
\section{Augmented IPW  for cross-sectional data}
\label{app: aipw}
Given our proposed heterogeneous treatment effect estimator, one direct way to construct an average treatment effect estimator is by building on results for doubly robust
estimation as developed in \citet{chernozhukov2016locally}. In order to do so, we first note that
the average treatment parameter can be written as a weighted average of outcomes using
inverse-probability-style weights as follows: $\tau = \EE{\gamma(Z_i) Y_i}$ with
\begin{equation}
	\label{eq:gamma}
	\gamma(Z_i)\ =\ \gamma(X_i,\ S_i,\ T_i)\ =\ \frac{T_i S_i}{e_{1,1}(X_i)} - \frac{(1-T_i) S_i}{e_{0,1}(X_i)} - \frac{T_i (1-S_i)}{e_{1,0}(X_i)}  + \frac{(1-T_i) (1-S_i)}{e_{0,0}(X_i)}.
\end{equation}Recall $g(z) = \EE{Y_i \cond X_i = x, \, S_i = s, \, T_i = t}$. The result of \citet{chernozhukov2016locally} implies that if we obtain good estimates both of $\hat{g}$ and the inverse-probability-style weights outlined above, then we can
obtain semiparametrically efficient estimates of $\tau$ using the doubly robust form such as in Augmented Inverse Propensity Weighitng (AIPW) \citep{robins1}. We will refer to this algorithm as \textbf{AIPW}, which takes on the following steps:

\begin{enumerate}
	\item Following the cross-fitting steps 2 to 4 of Algorithm \ref{alg:TR}, estimate the nuisance parameters $\hat{m}^{-{\mathcal{I}_{k}}}(x)$, $\hat{\varsigma}^{-{\mathcal{I}_{k}}}(x)$, $\hat{\nu}^{-{\mathcal{I}_{k}}}(x))$, $\hat{A}^{-{\mathcal{I}_{k}}}(z)$, $\hat{B}^{-{\mathcal{I}_{k}}}(z)$, $\hat{C}^{-{\mathcal{I}_{k}}}(z)$ using data not in the $\mathcal{I}_k$, where $\mathcal{I}_1$, ..., $\mathcal{I}_K$ are the $K$ folds of the data. In the same way, fit nonparametric regressions for the propensities $\hat{e}_{1,1}^{-{\mathcal{I}_{k}}}(x)$, $\hat{e}_{1,0}^{-{\mathcal{I}_{k}}}(x)$, $\hat{e}_{0,1}^{-{\mathcal{I}_{k}}}(x)$ and $\hat{e}_{0,0}^{-{\mathcal{I}_{k}}}(x)$.
	\item Run step 3 of Algorithm \ref{alg:h_did} to obtain cross-fitted point estimates $\hat{\tau}(X_j)=\hat{\tau}^{-\scaleto{\mathcal{I}_k}{5pt}}(X_j)$, for each $j \in \mathcal{I}_k$.
	\item For each $j\in \mathcal{I}_k$, produce the cross-fitted point estimates:
	\begin{equation}
		\begin{split}
			\hat{g}( Z_j) &= \hat{m}^{-{\mathcal{I}_{k}}}(X_j) + \hat{A}^{-{\mathcal{I}_{k}}}(Z_j)\hat{\nu}^{-{\mathcal{I}_{k}}}(X_j) \\
			&\ \ \ \ \ + \hat{B}^{-{\mathcal{I}_{k}}}(Z_j)\hat{\varsigma}^{-{\mathcal{I}_{k}}}(X_j) + \hat{C}^{-{\mathcal{I}_{k}}}(Z_j)\hat{\tau}^{-\scaleto{\mathcal{I}_k}{5pt}}(X_j)
		\end{split}
	\end{equation}
	and
	\begin{equation}\label{eq:aipw_gamma}
		\hat{\gamma}(Z_j) = \frac{T_j S_j}{\hat{e}^{-{\mathcal{I}_{j}}}_{1,1}(X_j)} - \frac{(1-T_j) S_j}{\hat{e}^{-{\mathcal{I}_{j}}}_{0,1}(X_j)} - \frac{T_j (1-S_j)}{\hat{e}^{-{\mathcal{I}_{j}}}_{1,0}(X_j)}  + \frac{(1-T_j) (1-S_j)}{\hat{e}^{-{\mathcal{I}_{j}}}_{0,0}(X_j)}.
	\end{equation}
	
	\item Estimate the average treatment effect $\EE{\tau(x)}$ as:
	\begin{equation}
		\label{eq:DR}
		\htau_{DR} = \frac{1}{n} \sum_{i = 1}^n \p{\htau(X_i) + \hat{\gamma}(Z_i) \p{Y_i - \hat{g}(Z_i)}}.
	\end{equation}
	
\end{enumerate}

\citet{abadie2005semiparametric} and \citet{sant2018doubly} explored a special case of this approach in cases where
$(X_i, S_i)$ are independent from $T_i$. As a result, only the estimate of the single propensity $s(x)$ is needed to perform a similar estimation, as opposed to the quantity $\gamma(x,s,t)$.

While plug-in estimation with the doubly robust score admits for algorithmically simple estimation of the
average treatment effect, probabilities $e_{1,1}(x)$ etc. may get quite small, and so inverting even slightly inaccurate propensity
estimates may result in instability. 

\bibliographystyle{plainnat}
\bibliography{references}

\pagebreak

\renewcommand*{\thesection}{\arabic{section}}
\begin{center}
	\textbf{\LARGE Supplemental Materials}
\end{center}

\setcounter{section}{0}
\renewcommand*{\theHsection}{chX.\the\value{section}}

\setcounter{equation}{0}
\setcounter{figure}{0}
\setcounter{table}{0}
\setcounter{page}{1}
\makeatletter
\renewcommand{\theequation}{S\arabic{equation}}
\renewcommand{\thefigure}{S\arabic{figure}}
\renewcommand{\bibnumfmt}[1]{[S#1]}
\renewcommand{\citenumfont}[1]{S#1}
\section{Proof of Proposition \ref{prop:decomp}}\label{ap:decomp_pf}

Assuming parallel trends conditioning on covariates $X_i$ as in \eqref{eq:cond-parallel}, $Y_i$ can be written from \eqref{eq:true_form} as follows:

\begin{equation}\label{eq:basic}
\begin{split}
Y_i &= (1-T_i-S_i+T_iS_i)g(X_i,0,0) + (T_i - T_iS_i)g(X_i,0,1) \\
&\ \ \ \ \ \ \ + (S_i - T_iS_i)g(X_i,1,0) + T_iS_ig(X_i,1,1) + \epsilon_i
\end{split}
\end{equation}
where $g(x,\ s,\ t) = \EE{Y_i \cond X_i = x, S_i = s, T_i = t}$. We could then write the above as:
\begin{equation}\label{eq:original}
Y_i = g(X_i,0,0) + T_i \rho(X_i) + S_i\xi(X_i) + T_iS_i\tau(X_i) + \epsilon_i
\end{equation}
where $\rho$ and $\xi$ are defined as in \eqref{eq:conditional_eff}. Note that ultimately, we want a decomposition of the following form:
\begin{equation}\label{eq:all_data_0}
Y_i = D(Z_i)\cdot m(X_i) + A(Z_i)\cdot \nu(X_i) + B(Z_i)\cdot \varsigma(X_i) + C(Z_i)\cdot \tau(X_i) + \epsilon_i
\end{equation}
we thus seek coefficients $A,B,C$ and $D$. Note that we have the following expressions:
\begin{align*}
m(X_i) &= e_{0,0}(X_i) g(X_i,0,0) + e_{0,1}(X_i) g(X_i,0,1) + e_{1,0}(X_i) g(X_i,1,0) + e_{1,1}(X_i) g(X_i,1,1) \\
\nu(X_i) &= \frac{e_{1,1}(X_i)}{t(X_i)} g(X_i,1,1) + \frac{e_{0,1}(X_i)}{t(X_i)} g(X_i,0,1) \\
&\ \ \ \ - \frac{e_{1,0}(X_i)}{(1-t(X_i))} g(X_i,1,0) - \frac{e_{0,0}(X_i)}{(1-t(X_i))} g(X_i,0,0) \\
\varsigma(X_i) &= \frac{e_{1,1}(X_i)}{s(X_i)} g(X_i,1,1) + \frac{e_{1,0}(X_i)}{s(X_i)} g(X_i,1,0) \\
&\ \ \ \ - \frac{e_{0,1}(X_i)}{(1-s(X_i))} g(X_i,1,0) - \frac{e_{0,0}(X_i)}{(1-s(X_i))} g(X_i,0,0) \\
\tau(X_i) &= g(X_i,0,0) - g(X_i,0,1) - g(X_i,1,0) + g(X_i,1,1).
\end{align*}
Equating the coefficients from (\ref{eq:all_data_0}) and (\ref{eq:basic}), we have that:
\begin{align*}
D(Z_i) e_{0,0}(X_i) - A(Z_i) \frac{e_{0,0}(X_i)}{(1-t(X_i))} - B(Z_i) \frac{e_{0,0}(X_i)}{(1-s(X_i))} + C(Z_i)  &= 1 - T_i - S_i + T_iS_i \\
D(Z_i) e_{0,1}(X_i) + A(Z_i) \frac{e_{0,1}(X_i)}{t(X_i)} - B(Z_i) \frac{e_{0,1}(X_i)}{(1-s(X_i))} - C(Z_i) &= T_i - T_iS_i \\
D(Z_i) e_{1,0}(X_i) - A(Z_i) \frac{e_{1,0}(X_i)}{(1-t(X_i))} + B(Z_i) \frac{e_{1,0}(X_i)}{s(X_i)} - C(Z_i) &= S_i - T_iS_i \\
D(Z_i) e_{1,1}(X_i) + A(Z_i) \frac{e_{1,1}(X_i)}{t(X_i)} + B(Z_i) \frac{e_{1,1}(X_i)}{s(X_i)} + C(Z_i) &= T_iS_i.
\end{align*}
Summing the equations, we get that $D \equiv 1$. We can then reformulate our objective as:
\begin{equation}\label{eq:all_data}
Y_i = m(X_i) + A(Z_i)\cdot \nu(X_i) + B(Z_i)\cdot \varsigma(X_i) + C(Z_i)\cdot \tau(X_i) + \epsilon_i
\end{equation}
which is the final form expressed in the proposition. We re-express the components as:
\begin{align*}
\nu(x) &= \rho(x) + \frac{e_{1,1}(x)}{t(x)} \tau(x) + \left(\frac{e_{1,1}(x)}{t(x)} - \frac{e_{1,0}(x)}{(1-t(x))}\right) \xi(x) \\
\varsigma(x) &= \xi(x) + \frac{e_{1,1}(x)}{t(x)} \tau(x) + \left(\frac{e_{1,1}(x)}{s(x)} - \frac{e_{0,1}(x)}{(1-s(x))}\right) \rho(x)\\
m(x) &=  g(x,0,0)+ (e_{1,1}(x) + e_{1,0}(x))\xi(x) + (e_{1,1}(x)+e_{0,1}(x)\rho(x) + e_{1,1}(x)\tau(x)
\end{align*}
equating coefficients of (\ref{eq:all_data}) and (\ref{eq:original}), we have:
\begin{align*}
\frac{e_{1,1}(X_i)}{t(X_i)} A(Z_i) + \frac{e_{1,1}(X_i)}{s(X_i)} B(Z_i) + C(Z_i) &= T_iS_i - e_{1,1}(X_i) \\
\left(\frac{e_{1,1}(X_i)}{t(X_i)} - \frac{e_{1,0}(X_i)}{(1-t(X_i))}\right)A(Z_i) + B(Z_i) &= S_i - s(X_i) \\
A(Z_i) + \left(\frac{e_{1,1}(X_i)}{s(X_i)} - \frac{e_{0,1}(X_i)}{(1-s(X_i))}\right) B(Z_i) &= T_i - t(X_i) \\
\end{align*}

Let us define $\Delta_{s}(x) = e_{1,1}(x)/s(x) - e_{0,1}(x)/(1-s(x))$ and $\Delta_{t}(x) = e_{1,1}(x)/t(x) - e_{1,0}(x)/(1-t(x))$. The second and third equations become:
\begin{align*}
\Delta_{t}(X_i) A(Z_i) + B(Z_i) &= S_i - s(X_i)\\
A(Z_i) + \Delta_{s}(X_i) B(Z_i) &= T_i - t(X_i)
\end{align*}
which gives us:
\begin{align*}
A(Z_i)&= \frac{1}{1-\Delta_{t}(X_i)\cdot \Delta_{s}(X_i)} \left(T_i - t(X_i) - \Delta_{s}(X_i) \cdot ( S_i - s(X_i)) \right) \\
B(Z_i) &= \frac{1}{1-\Delta_{t}(X_i)\cdot\Delta_{s}(X_i)}\left(S_i - s(X_i) - \Delta_{t}(X_i) \cdot (T_i - t(X_i)) \right) \\
C(Z_i) &= T_iS_i - e_{1,1}(X_i) - \frac{e_{1,1}(X_i)}{t(X_i)} A(Z_i) - \frac{e_{1,1}(X_i)}{s(X_i)} B(Z_i)
\end{align*}
To obtain the expressions in \eqref{eq:ABC}, we just have to get rid of the conditional expectations form above, and it is easy to check that we will obtain the desired forms. Now we check \eqref{eq:orthog_prop}. First, note that:
\begin{align*}
&\ \ \ \ \EE{A(Z_i)\cond X_i} \\
&= f(X_i)^{-1}\p{\EE{T_i\cond X_i} - t(X_i) - \frac{ \Delta(X_i) \p{\EE{S_i\cond X_i} - s(X_i)}}{s(X_i)(1 - s(X_i))}}
\end{align*}
where $f(X_i) = 1 - \frac{\Delta^2(X_i)}{s(X_i)(1 - s(X_i))t(X_i)(1 - t(X_i))}$. Because $\EE{T_i\cond X_i} = t(X_i)$ and $\EE{S_i\cond X_i} = s(X_i)$, the above expression evaluates to zero. Similarly, we can show $\EE{B(Z_i) \cond X_i} = \EE{C(Z_i) \cond X_i} = 0$. For terms such as $\EE{A(Z)\cond X, S}$, note that
\begin{align*}
&\ \ \ \ \EE{A(Z_i)\cond X_i, S_i = 1} \\
&= f(X_i)^{-1}\p{\EE{T_i\cond X_i, S_i = 1} - t(X_i) - \frac{ \Delta(X_i) }{s(X_i)}}.
\end{align*}
Because $\EE{T_i\cond X_i, S_i = 1} = e_{1,1}(X_i)/s(X_i)$, the above also evaluates to zero. Similarly, we can show that $\EE{A(Z_i)\cond X_i, S_i = 0}$ and $\EE{B(Z_i)\cond X_i, T_i} = 0$. For $\EE{C(Z)\cond X,S}$, note that:
\begin{align*}
\EE{B(Z_i)\cond X_i, S_i = 1} &= f(X_i)^{-1}\p{ 1-s(X_i) - \frac{\Delta(X_i)\p{\frac{e_{1,1}(X_i)}{s(X_i)} - t(X_i)}}{t(X_i)\p{1-t(X_i)}} } \\
&= 1 - s(X_i).
\end{align*}
Thus we have
\begin{align*}
\EE{C(Z_i)\cond X_i, S_i = 1} = \frac{e_{1,1}(X_i)}{s(X_i)}  - \frac{e_{1,1}}{s(X_i)}\p{1-s(X_i)} - e_{1,1}(X_i) = 0,
\end{align*}
and $\EE{C(Z_i)\cond X_i, S_i = 0} = 0$ and $\EE{C(Z_i)\cond X_i, T_i} = 0$ can be checked similarly. To check $C(Z_i)$ is uncorrelated with $A(Z_i)$ given $X_i$:
\begin{align*}
&\EE{C(Z_i)A(Z_i)\cond X_i} \\
&= \EE[S_i, T_i]{A(Z_i)\p{S_iT_i - e_{1,1}(X_i) - \frac{e_{1,1}(X_i)}{t(X_i)}A(Z_i) - \frac{e_{1,1}(X_i)}{s(X_i)}B(Z_i)}}\\
&=  \EE[S_i, T_i]{ \frac{1}{f(X_i)}\p{T_i-t(X_i)}\p{T_iS_i - e_{1,1}(X_i)} } \\
&\ \ \  - \EE[S_i, T_i]{ \frac{\Delta(X_i)}{f(X_i)s(X_i)(1-s(X_i))} (S_i - s(X_i))(T_iS_i-e_{1,1}(X_i))} \\
&\ \ \  - \mathbb{E}_{S_i,T_i}\Bigg[ \frac{e_{1,1}(X_i)}{t(X_i)f(X_i)^2}\cdot\bigg\{(T_i-t(X_i))^2 \\
&\ \ \ \ \ \ \ \ \ \ \ \ \ \ \ \ - 2\frac{\Delta(X_i)}{s(X_i)(1-s(X_i))}(T_i-t(X_i))(S_i-s(X_i)) \\
&\ \ \ \ \ \ \ \ \ \ \ \ \ \ \ \ + \frac{\Delta(X_i)^2}{s(X_i)^2(1-s(X_i))^2}(S_i-s(X_i))^2\bigg\}\Bigg]\\
&\ \ \  + \mathbb{E}_{S_i,T_i}\Bigg[\frac{e_{1,1}(X_i)}{s(X_i)f(X_i)^2}\cdot\bigg\{\frac{\Delta(X_i)}{t(X_i)(1-t(X_i))}(T_i-t(X_i))^2 \\
&\ \ \ \ \ \ \ \ \ \ \ \ \ \ \ \ - \frac{\Delta(X_i)^2}{s(X_i)(1-s(X_i))t(X_i)(1-t(X_i))}(T_i-t(X_i))(S_i - s(X_i)) \\
&\ \ \ \ \ \ \ \ \ \ \ \ \ \ \ \ - (T_i-t(X_i))(S_i - s(X_i)) + \frac{\Delta(X_i)}{s(X_i)(1-s(X_i))}(S_i-s(X_i))^2\bigg\}\Bigg]\\
&= \frac{e_{1,1}(X_i)}{f(X_i)}\p{(1-t(X_i)) - \frac{\Delta(X_i)}{s(X_i)} - (1-t(X_i)) + \frac{\Delta(X_i)}{s(X_i)}} \\
&= 0
\end{align*}
where the third equality follows by noting that 
\begin{align*}
&\EE{(T_i-t(X_i))^2\cond X_i} = t(X_i)(1-t(X_i),\\
&\EE{(S_i-s(X_i))^2\cond X_i} = s(X_i)(1-s(X_i)),\\
&\EE{(T_i - t(X_i))(S_i - s(X_i))\cond X_i} = \Delta(X_i).
\end{align*}
We can similarly check that $\EE{C(Z_i)B(Z_i)\cond X_i} = 0$.

\section{Proof of Theorem \ref{theo:const_tau}}\label{ap:const_tau_pf}
First we prove the statement $\sqrt{n}(\hat{\tau}_{TR} - \hat{\tau}^*) \xrightarrow{p}0$, where $\hat{\tau}_{TR} $ is the transformed regression estimator, and $\hat{\tau}^*$ is the transformed regression estimator with oracle nuisance parameters. From \eqref{eq:TR}, we see that
\begin{equation}
\sqrt{n}(\hat{\tau}_{TR} - \hat{\tau}^*) = \sqrt{n}\p{\sum_{k=1}^{K} \frac{|\mathcal{I}_k|}{n} \hat{\tau}^{-\scaleto{\mathcal{I}_k}{5pt}} - \hat{\tau}^*} = \sum_{k=1}^{K} \frac{|\mathcal{I}_k|}{n}\cdot \sqrt{n}\p{\hat{\tau}^{-\scaleto{\mathcal{I}_k}{5pt}} - \hat{\tau}^*}.
\end{equation}
Because each $|\mathcal{I}_k|/n$ is approximately $1/K$, which is fixed as $n$ grows, and the number of folds $K$ is also fixed, $\sqrt{n}(\hat{\tau}_{TR} - \hat{\tau}^*) \xrightarrow{p}0$ will follow if we show that $\sqrt{n}(\hat{\tau}^{-\scaleto{\mathcal{I}_k}{5pt}} - \hat{\tau}^*) \rightarrow_p 0$.

The proof consists of two steps. Firstly, we show that, if two nuisance parameters $\mu(x)$ and $\nu(x)$, estimated by $\hat{\mu}^{-\mathcal{I}_{k}}(x)$ and $\hat{\nu}^{-\mathcal{I}_{k}}(x)$ respectively, satisfy conditions 2, 3 and 4 from above, then the estimation of their product $\mu(x)\cdot\nu(x)$ by $\hat{\mu}^{-\mathcal{I}_{k}}(x)\cdot\hat{\nu}^{-\mathcal{I}_{k}}(x)$, and their sum $\mu(x) + \nu(x)$ by $\hat{\mu}^{-\mathcal{I}_{k}}(x) + \hat{\nu}^{-\mathcal{I}_{k}}(x)$, also satisfy conditions 2, 3, and 4. As a result, the estimate of $\hat{H}^{-\mathcal{I}_{k}}(x)$ and $\hat{C}^{-\mathcal{I}_{k}}(x)$, by sums and products of other nuisance parameters, such as $\hat{m}^{-\mathcal{I}_{k}}(x)$ and $\hat{e}^{-\mathcal{I}_{k}}_{1,1}(x)$, also satisfies the conditions listed above. Secondly, we show the desired result, assuming that $\hat{H}^{-\mathcal{I}_{k}}(x)$ and $\hat{C}^{-\mathcal{I}_{k}}(x)$ satisfy the conditions above.

We start with the first step. Assume that $\hat{\mu}^{-\mathcal{I}_{k}}(x)$ and $\hat{\mu}^{-\mathcal{I}_{k}}(x)$ satisfy conditions 2 to 4. We want to estimate $\mu(x)\cdot\nu(x)$ by $\hat{\mu}^{-\mathcal{I}_{k}}(x)\cdot\hat{\nu}^{-\mathcal{I}_{k}}(x)$, and $\mu(x) + \nu(x)$ by $\hat{\mu}^{-\mathcal{I}_{k}}(x) + \hat{\nu}^{-\mathcal{I}_{k}}(x)$. Consistency and boundedness comes easily from the consistency and boundedness of $\hat{\mu}^{-\mathcal{I}_{k}}(x)$ and $\hat{\mu}^{-\mathcal{I}_{k}}(x)$. Risk decay comes as follows:
\begin{equation*}
\begin{aligned}
&\mathbb{E}[(\hat{\mu}^{-\mathcal{I}_{k}}(x)\cdot\hat{\nu}^{-\mathcal{I}_{k}}(x) - \mu(x)\cdot\nu(x))^2] \\
={} &\mathbb{E}\left[\left(\frac{1}{2}(\hat{\mu} + \mu)(\hat{\nu} - \nu) + \frac{1}{2}(\hat{\nu} + \nu)(\hat{\mu} - \mu)\right)^2\right] \\
={} &\mathbb{E}\left[\left(\frac{1}{2}(\hat{\mu} + \mu)(\hat{\nu} - \nu)\right)^2\right] + \mathbb{E}\left[\left(\frac{1}{2}(\hat{\nu} + \nu)(\hat{\mu} - \mu)\right)^2\right] \\
&+\mathbb{E}\left[\left(\frac{1}{2}(\hat{\mu} + \mu)(\hat{\nu} - \nu)\right) \left(\frac{1}{2}(\hat{\nu} + \nu)(\hat{\mu} - \mu)\right)\right] \\
\leq{} &2\left(\mathbb{E}\left[\left(\frac{1}{2}(\hat{\mu} + \mu)(\hat{\nu} - \nu)\right)^2\right] + \mathbb{E}\left[\left(\frac{1}{2}(\hat{\nu} + \nu)(\hat{\mu} - \mu)\right)^2\right]\right) \\
\leq{} &M^2 \left(\mathbb{E}\left[(\hat{\mu} - \mu)^2\right]+\mathbb{E}\left[(\hat{\nu} - \nu)^2\right]\right) \\
={} &o\left(\frac{1}{\sqrt{n}}\right)
\end{aligned}
\end{equation*}
where we used the assumption that the nuisance parameters are all bounded by $M$. Risk decay for $\mu + \nu$ follows similarly. Recall we want to show the following:

\begin{equation}
\begin{split}
\enspace \hat{\tau}^{-\scaleto{\mathcal{I}_k}{5pt}} - \hat{\tau}^* &= \frac{\frac{1}{n}\sum_{j\in \mathcal{I}_k} \hat{H}^{-\mathcal{I}_{k}}(Z_j) \hat{C}^{-\mathcal{I}_{k}}(Z_j)}{\frac{1}{n}\sum_{j\in \mathcal{I}_k} \hat{C}^{-\mathcal{I}_{k}}(Z_j)^2} - \frac{\frac{1}{n}\sum_{j\in \mathcal{I}_k} H(Z_j)C(Z_j)}{\frac{1}{n}\sum_{j\in \mathcal{I}_k} C(Z_j)^2} \\
&= o_p\left(\frac{1}{\sqrt{n}}\right),
\end{split}
\end{equation}
which will follow if we can show that
\begin{equation}\label{eq:numerator}
\frac{1}{n}\sum_{j\in \mathcal{I}_k} \hat{H}^{-\mathcal{I}_{k}}(Z_j)\hat{C}^{-\mathcal{I}_{k}}(Z_j) - \frac{1}{n}\sum_{j\in \mathcal{I}_k} H(Z_j)C(Z_j)  = o_p(\frac{1}{\sqrt{n}})
\end{equation}
and
\begin{equation}\label{eq:denominator}\frac{1}{n}\sum_{j\in \mathcal{I}_k} \hat{C}^{-\mathcal{I}_{k}}(Z_j)^2 - \frac{1}{n}\sum_{j\in \mathcal{I}_k} C(Z_j)^2 = o_p(\frac{1}{\sqrt{n}})
\end{equation}
Because with these bounds, and noting the Taylor expansion of $f(x,y) = \frac{x}{y}$ at some point $x_0$ and $y_0 \neq 0$ is given by:
$$
f(x,y) = f(x_0, y_0) + \frac{1}{y_0}\cdot (x-x_0) - \frac{x_0}{y_0^2} \cdot (y-y_0) + O\left((x-x_0)^2 + (y-y_0)^2\right)
$$
we have that:
\begin{equation}
\begin{aligned}
\hat{\tau}_{\scaleto{\mathcal{I}_k}{5pt}} - \hat{\tau}^* ={} &\frac{\frac{1}{n}\sum_{j\in \mathcal{I}_k} \hat{H}^{-\mathcal{I}_{k}}(Z_j) \hat{C}^{-\mathcal{I}_{k}}(Z_j)}{\frac{1}{n}\sum_{j\in \mathcal{I}_k} \hat{C}^{-\mathcal{I}_{k}}(Z_j)^2} - \frac{\frac{1}{n}\sum_{j\in \mathcal{I}_k} H(Z_j)C(Z_j)}{\frac{1}{n}\sum_{j\in \mathcal{I}_k} C(Z_j)^2} \\
={} &f\left(\frac{1}{n}\sum_{j\in \mathcal{I}_k} \hat{H}^{-\mathcal{I}_{k}}(Z_j) \hat{C}^{-\mathcal{I}_{k}}(Z_j), \frac{1}{n}\sum_{j\in \mathcal{I}_k} \hat{C}^{\mathcal{I}_{k}}(X_j)^2\right) \\
&- f\left(\frac{1}{n}\sum_{j\in \mathcal{I}_k} H(Z_j)C(Z_j), \frac{1}{n}\sum_{j\in \mathcal{I}_k} C(Z_j)^2\right)\\
={} &\frac{1}{\frac{1}{n}\sum_{j\in \mathcal{I}_k} C(Z_j)^2}\cdot \left(\frac{1}{n}\sum_{j\in \mathcal{I}_k} \hat{H}^{-\mathcal{I}_{k}}(Z_j) \hat{C}^{-\mathcal{I}_{k}}(Z_j) - \frac{1}{n}\sum_{j\in \mathcal{I}_k} H(Z_j)C(Z_j) \right)\\
 &- \frac{\frac{1}{n}\sum_{j\in \mathcal{I}_k} H(Z_j)C(Z_j)}{\left(\frac{1}{n}\sum_{j\in \mathcal{I}_k} C(Z_j)^2\right)^2}\cdot \left(\frac{1}{n}\sum_{j\in \mathcal{I}_k} \hat{C}^{\mathcal{I}_{k}}(X_j)^2 - \frac{1}{n}\sum_{j\in \mathcal{I}_k} C(Z_j)^2 \right) + o_p\left(\frac{1}{n}\right) \label{quantities}\\
={} &o_p\left(\frac{1}{\sqrt{n}}\right) 
\end{aligned}
\end{equation}
where Slutsky's theorem is used in the last equality. Let us first check \eqref{eq:numerator}. Note that:
\begin{align}
&\frac{1}{n}\sum_{j\in \mathcal{I}_k} \hat{H}^{-\mathcal{I}_{k}}(Z_j)\hat{C}^{-\mathcal{I}_{k}}(Z_j) - \frac{1}{n}\sum_{j\in \mathcal{I}_k} H(Z_j)C(Z_j) \\
&= \frac{1}{n}\sum_{j\in \mathcal{I}_k} H(Z_j)\p{ \hat{C}^{-\mathcal{I}_{k}}(Z_j) - C(Z_j) }\label{eq:mean0_1}\\
&\ \ \ \ +\frac{1}{n}\sum_{j\in \mathcal{I}_k} C(Z_j)\p{\hat{H}^{-\mathcal{I}_{k}}(Z_j) - H(Z_j) }\label{eq:mean0_2}\\
&\ \ \ \ +\frac{1}{n}\sum_{j\in \mathcal{I}_k} \p{\hat{H}^{-\mathcal{I}_{k}}(Z_j)-H(Z_j)}\p{\hat{C}^{-\mathcal{I}_{k}}(Z_j)-C(Z_j)}\label{eq:cs1}.
\end{align}
We check that the three terms above are small. 

For \eqref{eq:mean0_1}, we first check that $H(Z_j)\p{ \hat{C}^{-\mathcal{I}_{k}}(Z_j) - C(Z_j) }$ has mean zero. Note that
\begin{align*}
&\EE{H(Z_j)\p{ \hat{C}^{-\mathcal{I}_{k}}(Z_j) - C(Z_j) }}\\
&= \mathbb{E}\Bigg[\p{\tau(X_i)C(Z_i)+\epsilon_i}\p{e_{1,1}(X_i) + J(X_i)A(Z_i) + K(X_i)B(Z_i) }\\
&\ \ \ \ \ \  - \p{\tau(X_i)C(Z_i)+\epsilon_i}\p{\hat{e}^{-\mathcal{I}_{k}}_{1,1}(X_i)+\hat{J}^{-\mathcal{I}_{k}}(X_i)\hat{A}^{-\mathcal{I}_{k}}(Z_i) + \hat{K}^{-\mathcal{I}_{k}}(X_i)\hat{B}^{-\mathcal{I}_{k}}(Z_i)}\Bigg]\\
&= \EE{\tau(X_i)\hat{J}^{-\mathcal{I}_{k}}(X_i)\hat{A}^{-\mathcal{I}_{k}}(Z_i)C(Z_i)} + \EE{\tau(X_i)\hat{K}^{-\mathcal{I}_{k}}(X_i)\hat{B}^{-\mathcal{I}_{k}}(Z_i)C(Z_i)},
\end{align*}
where $J(X_i) = e_{1,1}(X_i)/t(X_i)$ and $K(X_i) = e_{1,1}(X_i)/s(X_i)$ and we similarly define the cross-fitted analogue $\hat{J}^{-\ii_k}(\cdot)$ and $\hat{K}^{-\ii_k}(\cdot)$. Many terms above vanished because of \eqref{eq:orthog_prop}. We will just show that 
\begin{equation}\label{eq:showing_mean0}
\EE{\tau(X_i)\hat{J}^{-\mathcal{I}_{k}}(X_i)\hat{A}^{-\mathcal{I}_{k}}(Z_i)C(Z_i)}=0,
\end{equation}
and the argument for $\EE{\tau(X_i)\hat{K}^{-\mathcal{I}_{k}}(X_i)\hat{B}^{-\mathcal{I}_{k}}(Z_i)C(Z_i)}$ is identical. Note that 
\begin{equation}\tau(X_i)\hat{J}^{-\mathcal{I}_{k}}(X_i)\hat{A}^{-\mathcal{I}_{k}}(Z_i) = \hat{L}^{-\mathcal{I}_{k}}(X_i) + \hat{N}^{-\mathcal{I}_{k}}(X_i)T_i + \hat{P}^{-\mathcal{I}_{k}}(X_i)S_i,
\end{equation}
where
\begin{align*}
&\hat{L}^{-\mathcal{I}_{k}}(X_i) = \tau(X_i)\hat{J}^{-\mathcal{I}_{k}}(X_i)\cdot\frac{1}{\hat{f}^{-\mathcal{I}_{k}}(X_i)}\p{\frac{\hat{\Delta}^{-\mathcal{I}_{k}}(X_i)}{1-\hat{s}^{-\mathcal{I}_{k}}(X_i)} - \hat{t}^{-\mathcal{I}_{k}}(X_i)},\\
&\hat{N}^{-\mathcal{I}_{k}}(X_i)=\tau(X_i)\hat{J}^{-\mathcal{I}_{k}}(X_i)\cdot\frac{1}{\hat{f}^{-\mathcal{I}_{k}}(X_i)},\\
&\hat{P}^{-\mathcal{I}_{k}}(X_i)= -\tau(X_i)\hat{J}^{-\mathcal{I}_{k}}(X_i)\cdot\frac{1}{\hat{f}^{-\mathcal{I}_{k}}(X_i)}\frac{\hat{\Delta}^{-\mathcal{I}_{k}}(X_i)}{\hat{s}^{-\mathcal{I}_{k}}(X_i)\p{1-\hat{s}^{-\mathcal{I}_{k}}(X_i)}}.
\end{align*}
Thus, again by \eqref{eq:orthog_prop}, we have
\begin{align*}
\EE{\hat{L}^{-\mathcal{I}_{k}}(X_i)C(Z_i)} &= \EE{\EE{\hat{L}^{-\mathcal{I}_{k}}(X_i)C(Z_i)\cond -\mathcal{I}_{k}}}\\
&=\EE{\EE{\hat{L}^{-\mathcal{I}_{k}}(X_i)\EE{C(Z_i)\cond X_i}\cond -\mathcal{I}_{k}}}\\
&= 0
\end{align*}
where the first line conditions on the data not in fold $\mathcal{I}_{k}$. Similarly, we have that
$$
\EE{\hat{N}^{-\mathcal{I}_{k}}(X_i)T_iC(Z_i)} = \EE{\hat{P}^{-\mathcal{I}_{k}}(X_i)S_iC(Z_i)} = 0.
$$
Hence we know that $\EE{\tau(X_i)\hat{J}^{-\mathcal{I}_{k}}(X_i)\hat{A}^{-\mathcal{I}_{k}}(Z_i)C(Z_i)}$, and subsequently \newline$\EE{H(Z_j)\p{ \hat{C}^{-\mathcal{I}_{k}}(Z_j) - C(Z_j) }}$, are all zero. Now we are ready to show that \eqref{eq:mean0_1} is $o_p\p{\frac{1}{\sqrt{n}}}$:
\begin{align}
&\ \ \ \ \EE{\p{\frac{1}{n}\sum_{j\in \mathcal{I}_k} H(Z_j)\p{ \hat{C}^{-\mathcal{I}_{k}}(Z_j) - C(Z_j) }}^2}\\
&= \frac{\abs{\ii_k}}{n}\frac{1}{n}\EE{ H(Z_j)^2\p{ \hat{C}^{-\mathcal{I}_{k}}(Z_j) - C(Z_j) }^2}\\
&\leq \frac{\abs{\ii_k}}{n} \frac{M^2}{n}\EE{\p{ \hat{C}^{-\mathcal{I}_{k}}(Z_j) - C(Z_j) }^2}\\
&= o\p{\frac{1}{n^{3/2}}}
\end{align}
where the first equality is precisely because all terms of the form $H(Z_j)\p{ \hat{C}^{-\mathcal{I}_{k}}(Z_j) - C(Z_j) }$ have mean zero, and the bound comes from the risk decay assumption. To show \eqref{eq:mean0_2} is $o_p\p{\frac{1}{\sqrt{n}}}$, the argument is exactly the same as the one we gave above: the only difference being we have to check that the term $ C(Z_j)\p{\hat{H}^{-\mathcal{I}_{k}}(Z_j) - H(Z_j) }$ has mean zero. Note that:
\begin{align*}
&\EE{C(Z_j)\p{ \hat{H}^{-\mathcal{I}_{k}}(Z_j) - H(Z_j) }}\\
&= \mathbb{E}\Bigg[C(Z_i)\p{ m(X_i) + \nu(X_i)A(Z_i) + \varsigma(X_i)B(Z_i) }\\
&\ \ \ \  - C(Z_i)\p{\hat{m}^{-\mathcal{I}_{k}}_{1,1}(X_i)+\hat{\nu}^{-\mathcal{I}_{k}}(X_i)\hat{A}^{-\mathcal{I}_{k}}(Z_i) + \hat{\varsigma}^{-\mathcal{I}_{k}}(X_i)\hat{B}^{-\mathcal{I}_{k}}(Z_i)}\Bigg]\\
&= \EE{-\hat{\nu}^{-\mathcal{I}_{k}}(X_i)\hat{A}^{-\mathcal{I}_{k}}(Z_i)C(Z_i)} + \EE{-\hat{\varsigma}^{-\mathcal{I}_{k}}(X_i)\hat{B}^{-\mathcal{I}_{k}}(Z_i)C(Z_i)},
\end{align*}
where terms vanish in the last equality because of \eqref{eq:orthog_prop}. Again we just have to show that $\hat{\nu}^{-\mathcal{I}_{k}}(X_i)\hat{A}^{-\mathcal{I}_{k}}(Z_i)C(Z_i)$ and $\hat{\varsigma}^{-\mathcal{I}_{k}}(X_i)\hat{B}^{-\mathcal{I}_{k}}(Z_i)C(Z_i)$ have mean zero, which follows exactly the same argument as we used for \eqref{eq:showing_mean0}. Thus \eqref{eq:mean0_2} is also $o_p\p{\frac{1}{\sqrt{n}}}$. We only have to check that \eqref{eq:cs1} is $o_p\p{\frac{1}{\sqrt{n}}}$, which follows by Cauchy-Schwarz and the risk decay assumption:
\begin{align*}
&\  \ \ \ \frac{1}{n}\sum_{j\in \mathcal{I}_k} \p{\hat{H}^{-\mathcal{I}_{k}}(Z_j)-H(Z_j)}\p{\hat{C}^{-\mathcal{I}_{k}}(Z_j)-C(Z_j)}\\
&\leq \p{\frac{1}{n}\sum_{j\in \mathcal{I}_k} \p{\hat{H}^{-\mathcal{I}_{k}}(Z_j)-H(Z_j)}^2}^{\frac{1}{2}}\p{\frac{1}{n}\sum_{j\in \mathcal{I}_k} \p{\hat{C}^{-\mathcal{I}_{k}}(Z_j)-C(Z_j)}^2}^{\frac{1}{2}}\\
&= o_p\p{\frac{1}{\sqrt{n}}}.
\end{align*}
Thus we have finished checking \eqref{eq:numerator}. We proceed to check \eqref{eq:denominator}. Note that:
\begin{align}
&\ \ \ \  \frac{1}{n}\sum_{j\in \mathcal{I}_k} \hat{C}^{-\mathcal{I}_{k}}(Z_j)^2 - \frac{1}{n}\sum_{j\in \mathcal{I}_k} C(Z_j)^2\\
&= 2\cdot\frac{1}{n}\sum_{j\in \mathcal{I}_k} C(Z_j)\p{\hat{C}^{-\mathcal{I}_{k}}(Z_j)- C(Z_j)}\label{eq:mean0_3}\\
&\ \ \ \ \  + \frac{1}{n}\sum_{j\in \mathcal{I}_k} \p{\hat{C}^{-\mathcal{I}_{k}}(Z_j) -  C(Z_j)}^2.
\end{align}
The second term above is immediately $o_p\p{\frac{1}{\sqrt{n}}}$ because of the risk decay assumption. Thus we only have to check that \eqref{eq:mean0_3} is $o_p\p{\frac{1}{\sqrt{n}}}$, which will follow exactly the same argument for \eqref{eq:mean0_1}, once we show that $C(Z_j)\p{\hat{C}^{-\mathcal{I}_{k}}(Z_j)- C(Z_j)}$ has mean zero. Note then
\begin{align*}
&\EE{C(Z_j)\p{ \hat{C}^{-\mathcal{I}_{k}}(Z_j) - C(Z_j) }}\\
&= \mathbb{E}\Bigg[-C(Z_i)\p{e_{1,1}(X_i) + J(X_i)A(Z_i) + K(X_i)B(Z_i) }\\
&\ \ \ \  + C(Z_i)\p{\hat{e}^{-\mathcal{I}_{k}}_{1,1}(X_i)+\hat{J}^{-\mathcal{I}_{k}}(X_i)\hat{A}^{-\mathcal{I}_{k}}(Z_i) + \hat{K}^{-\mathcal{I}_{k}}(X_i)\hat{B}^{-\mathcal{I}_{k}}(Z_i)}\Bigg]\\
&= \EE{\hat{J}^{-\mathcal{I}_{k}}(X_i)\hat{A}^{-\mathcal{I}_{k}}(Z_i)C(Z_i)} + \EE{\hat{K}^{-\mathcal{I}_{k}}(X_i)\hat{B}^{-\mathcal{I}_{k}}(Z_i)C(Z_i)},
\end{align*}
where terms vanish in the last equality because of \eqref{eq:orthog_prop}. Following the same argument for showing \eqref{eq:showing_mean0}, we can show that the two terms above are both zero. Thus \newline$C(Z_j)\p{\hat{C}^{-\mathcal{I}_{k}}(Z_j)- C(Z_j)}$ indeed has mean zero and so is $o_p\p{\frac{1}{\sqrt{n}}}$, so we have finished showing $\sqrt{n}(\hat{\tau}_{TR} - \hat{\tau}^*) \xrightarrow{p}0$.

Now we show that $\sqrt{n}(\hat{\tau}^* - \tau)\xrightarrow{d} \nn(0,V_{TR})$, where $V_{TR} = \frac{\EE{\sigma^2(z) C^2(z)}}{\EE{C^2(z)}^2}$. Note that
\begin{align}
\sqrt{n}\p{\hat{\tau}^* -\tau}&= \frac{\sqrt{n}\p{\frac{1}{n}\sum_{i=1}^n H(Z_i)C(Z_i) - \p{\frac{1}{n}\sum_{i=1}^n C(Z_i)^2}\tau}}{\frac{1}{n}\sum_{i=1}^n C(Z_i)^2}\\
&= \frac{\sqrt{n}\p{\frac{1}{n}\sum_{i=1}^n \epsilon_iC(Z_i)}}{\frac{1}{n}\sum_{i=1}^n C(Z_i)^2}\\ \label{eq:slutsky1}
&\xrightarrow{d} \frac{1}{\EE{C(z)^2}}\cdot\nn(0, \EE{\sigma(z)^2C(z)^2})\\
&\overset{d}{=} \nn\p{0,\frac{\EE{\sigma^2(z) C^2(z)}}{\EE{C^2(z)}^2}}
\end{align}
where \eqref{eq:slutsky1} use Slutsky's theorem and the central limit theorem. The desired result then follows.

\section{Proof of Proposition \ref{prop:weight}}\label{ap:weighted_pf}
The proof of Proposition \ref{prop:weight} is essentially the same as that of Theorem \ref{theo:const_tau}: by exactly the same argument, we know that $\sqrt{n}(\hat{\tau}_{TR} - \hat{\tau}^*)\xrightarrow{p}0$, where $\hat{\tau}^*$ is the transformed regression estimator with oracle nuisance parameters. Then we just have to show $\sqrt{n}\p{\hat{\tau}^* - \bar{t}} \xrightarrow{d} \nn(0, V_{TR})$, where $V_{TR} = \frac{\EE{\sigma^2(z) C^2(z)}}{\EE{C^2(z)}^2}$, which follows from the following calculation:
\begin{align}
\sqrt{n}\p{\hat{\tau}^* -\bar{\tau}}&= \frac{\sqrt{n}\p{\frac{1}{n}\sum_{i=1}^n H(Z_i)C(Z_i) - \p{\frac{1}{n}\sum_{i=1}^n C(Z_i)^2}\bar{\tau}}}{\frac{1}{n}\sum_{i=1}^n C(Z_i)^2}\\
&= \frac{\sqrt{n}\p{\frac{1}{n} \sum_{i=1}^n \epsilon_iC(Z_i)}}{\frac{1}{n}\sum_{i=1}^n C(Z_i)^2}\\\label{eq:slutsky4}
&\xrightarrow{d} \frac{1}{\EE{C(z)^2}}\cdot\nn(0, \EE{\sigma(z)^2C(z)^2})\\
&\overset{d}{=} \nn\p{0,\frac{\EE{\sigma^2(z) C^2(z)}}{\EE{C^2(z)}^2}}
\end{align}
where \eqref{eq:slutsky4} uses Slutsky's theorem and the central limit theorem.

\section{Proof of Theorem \ref{theo:hte-rate}}

	Following notation in \citet{foster2019orthogonal}, we first define directional derivatives: we define $D_f(F)(f)[h] = \frac{d}{dt} F(f+th)\vert_{t=0}$ for a pair fo functions $f, h$. We define \newline$D_f^k(F)(f)[h_1, \ldots, h_k] = \frac{\partial^k}{\partial t_1\ldots \partial t_k} F(f+t_1h_1 + \ldots + t_kh_k)\vert_{t_1=\ldots=t_k=0}$. Next, we check Assumptions 1-4 stated in \citet{foster2019orthogonal}. 
	First, we check Assumption 2 on the first order optimality. 
	\begin{align*}
		&	D_\tau L(\tau, \mu)[\tau' - \tau] \\
		&= -2 \EE{\p{Y-m(X) - A(Z) \nu(X) - B(Z) \varsigma(X) -  C(Z) \tau(Z)} C(Z)(\tau'(X)-\tau(X))}\\
		&=-2 \EE{\EE{\p{Y-m(X) - A(Z) \nu(X) - B(Z) \varsigma(X) -  C(Z) \tau(Z)} C(Z) (\tau'(X)- \tau(X))\cond Z}}\\
		&= -2\EE{\EE{\epsilon C(Z)(\tau'(X)-\tau(X))\cond Z}} = 0
	\end{align*}
	due to the fact that $\EE{\epsilon \cond Z} =0$ from Proposition \ref{prop:decomp}.
	
	Next, we check Assumption 1. Note that we have to check it with directional derivative with respect to all nuisance components $m(x), \nu(x), \varsigma(x),  s(x), t(x), \Delta(x)$. By symmetry, we only need to check it with $m(x), \nu(x), s(x), \Delta(x)$. 
	\begin{align*}
		&D_m D_\tau L(\tau, \{m(x), \nu(x), \varsigma(x),  s(x), t(x), \Delta(x)\})[\tau' - \tau, m'-m] \\
		&=  -2 d \mathbbm{E}\left[(Y-(m(X)+ t_m(m'(X)-m(X)) - A(Z) \nu(X) - B(Z) \varsigma(X)  \right.\\
		&\left.\ \ \ \ 	-  C(Z) (\tau(X) + t_\tau(\tau'(X) - \tau(X)) )) C(Z)(\tau'(X)-\tau(X))\right]/dt_m \vert_{t_m=0, t_\tau=0}\\
		&= -2\EE{(m'(X)-m(X))C(Z) (\tau'(X) - \tau(X))} \\
		&= -2\EE{\EE{(m'(X)-m(X))C(Z) (\tau'(X) - \tau(X))\cond X}} \\
		&= -2\EE{(m'(X)-m(X)) (\tau'(X) - \tau(X))\EE{C(Z)\cond X}} \\
		&=0
	\end{align*}
	where the last equality follows from $\EE{C(Z)\cond X}=0$ as in Proposition \ref{prop:decomp}.
	
	Next, we check 
	\begin{align*}
		&D_\nu D_\tau L(\tau, \{m(x), \nu(x), \varsigma(x),  s(x), t(x), \Delta(x)\})[\tau' - \tau, \nu'-\nu] \\
		&=  -2 d \mathbbm{E}\left[(Y-m(X) - A(Z)( \nu(X) + t_\nu(\nu'(X)-\nu(X)))- B(Z) \varsigma(X)  \right.\\
		&\left.\ \ \ \ 	-  C(Z) (\tau(X) + t_\tau(\tau'(X) - \tau(X)) )) C(Z)(\tau'(X)-\tau(X))\right]/dt_\nu \vert_{t_\nu=0, t_\tau=0}\\
		&= -2\EE{A(Z)(\nu'(X)-\nu(X))C(Z) (\tau'(X) - \tau(X))} \\
		&= -2\EE{\EE{A(Z)(\nu'(X)-\nu(X))C(Z) (\tau'(X) - \tau(X))\cond X} }\\
		&= -2\EE{(\nu'(X)-\nu(X))(\tau'(X) - \tau(X))\EE{A(Z)C(Z) \cond X} }\\
		&=0
	\end{align*}
	where the last equality follows from $\EE{A(Z)C(Z)\cond X}=0$ as in Proposition \ref{prop:decomp}.
	
	Before we continue, we define  
		\begin{align*}
		A_{t_{\Delta}}(Z_i) = &\p{1 - \frac{(\Delta+t_\Delta (\Delta' - \Delta))^2(X_i)}{s(X_i)(1 - s(X_i))t(X_i)(1 - t(X_i))}}^{-1} \\ 
		&\p{T_i - t(X_i) - \frac{ (\Delta+t_\Delta (\Delta' - \Delta)) \p{S_i - s(X_i)}}{s(X_i)(1 - s(X_i))}}
	\end{align*}
	and similarly, we define $B_{t_\Delta}(Z_i), C_{t_\Delta}(Z_i)$.
	We further define 
	\begin{align*}
		M_{t_\Delta, t_\tau}(Z) = Y-m(X) - A_{t_{\Delta}}(Z)\nu(X)- B_{t_{\Delta}}(Z) \varsigma(X)  
		-  C_{t_{\Delta}}(Z) (\tau(X) + t_\tau(\tau'(X) - \tau(X)) )
	\end{align*}
	and
	\begin{align*}
		M(Z)= Y-m(X) - A(Z)\nu(X)- B(Z) \varsigma(X)  
		-  C(Z) \tau(X).
	\end{align*}

	We check 
	\begin{align*}
		&D_\Delta D_\tau L(\tau, \{m(x), \nu(x), \varsigma(x),  s(x), t(x), \Delta(x)\})[\tau' - \tau, \Delta'-\Delta] \\
		&=  -2 d \mathbbm{E}[M_{t_\Delta, t_\tau} (Z)C_{t_\Delta}(Z)(\tau'(X) - \tau(X))] / dt_\Delta\vert_{t_\Delta=0, t_\tau=0}\\
		&= -2 \EE{C(Z)(\tau'(X)-\tau(X)) \frac{\partial M_{t_\Delta, 0}(Z)}{\partial t_\Delta}\bigg|_{t_\Delta=0}} \\
		&\ \ \ \ \ -2 \EE{M(Z)(\tau'(X)-\tau(X)) \frac{\partial C_{t_\Delta}(Z)}{\partial t_\Delta}\bigg|_{t_\Delta=0}}.
	\end{align*}
	For the second term, we notice that $\EE{M(Z) \cond Z} = 0$ from Proposition \ref{prop:decomp}, and so
	\begin{align*}
		&-2 \EE{M(Z)(\tau'(X)-\tau(X)) \frac{\partial C_{t_\Delta}(Z)}{\partial t_\Delta}\bigg|_{t_\Delta=0}} \\
		&= -2 \EE{\EE{M(Z)(\tau'(X)-\tau(X)) \frac{\partial C_{t_\Delta}(Z)}{\partial t_\Delta}\bigg|_{t_\Delta=0}\cond Z}} 
		\\
		&= -2 \EE{(\tau'(X)-\tau(X)) \frac{\partial C_{t_\Delta}(Z)}{\partial t_\Delta}\bigg|_{t_\Delta=0}\EE{M(Z)\cond Z}}\\
		&=0.
	\end{align*}
	For the first term, 
	\begin{align*}
		&-2 \EE{C(Z)(\tau'(X)-\tau(X)) \frac{\partial M_{t_\Delta, 0}(Z)}{\partial t_\Delta}\bigg|_{t_\Delta=0}} \\
		&= 2\mathbb{E}\bigg[C(Z)(\tau'(X)-\tau(X)) \bigg(\frac{\partial A_{t_\Delta}(Z) \nu(X)}{\partial t_\Delta}\bigg|_{t_\Delta=0} +\frac{\partial B_{t_\Delta}(Z) \varsigma(X)}{\partial t_\Delta}\bigg|_{t_\Delta=0}  \\
				&\ \ \ + \frac{\partial C_{t_\Delta}(Z) \tau(X)}{\partial t_\Delta}\bigg|_{t_\Delta=0} \bigg)\bigg].
	\end{align*}	
	First, we note that 
	\begin{align*}
		&\frac{\partial A_{t_\Delta}(Z) \nu(X)}{\partial t_\Delta}\bigg|_{t_\Delta=0} \\
		&=\frac{\partial \p{1 - \frac{(\Delta+t_\Delta (\Delta' - \Delta))^2(X)}{s(X)(1 - s(X))t(X)(1 - t(X))}}^{-1}}{\partial t_\Delta}\bigg|_{t_\Delta=0} \p{T - t(X) - \frac{ \Delta(X) \p{S - s(X)}}{s(X)(1 - s(X))}}\nu(X) \\
		&\ \ \ \ +  \p{1 - \frac{\Delta^2(X)}{s(X)(1 - s(X))t(X)(1 - t(X))}}^{-1} \\
		&\ \ \ \ \frac{\partial \p{T - t(X) - \frac{ (\Delta(X)+t_\Delta(\Delta'(X)-\Delta)) \p{S - s(X)}}{s(X)(1 - s(X))}}}{\partial t_\Delta}\bigg|_{t_\Delta=0}\nu(X)\\
		&= f_1(X) A(Z) + f_2(X) f_3(X, S)
	\end{align*}
	for some function $f_1, f_2, f_3$. The upshot is that
	\begin{align*}
		&	2\EE{C(Z)(\tau'(X)-\tau(X)) \p{\frac{\partial A_{t_\Delta}(Z) \nu(X)}{\partial t_\Delta}\bigg|_{t_\Delta=0}}} \\
		&=2\EE{C(Z)(\tau'(X)- \tau(X))f_1(X)A(Z)} + 2\EE{C(Z)(\tau'(X)- \tau(X))f_2(X)f_3(X,S)} \\
		&=2\EE{\EE{C(Z)(\tau'(X)- \tau(X))f_1(X)A(Z)\cond X}} \\
		&\ \ \ \ \ + 2\EE{\EE{C(Z)(\tau'(X)- \tau(X))f_2(X)f_3(X,S)\cond X, S}}\\
		&= 2\EE{(\tau'(X)- \tau(X))f_1(X)\EE{C(Z)A(Z)\cond X}} \\
		&\ \ \ \ \ + 2\EE{(\tau'(X)- \tau(X))f_2(X)f_3(X,S)\EE{C(Z)\cond X, S}}\\
		&= 0+0 = 0
	\end{align*}
	where the second to the last equality follows from $\EE{C(Z) \cond X,S}=0$ and \newline $\EE{C(Z)A(Z)\cond X}=0$ from Proposition \ref{prop:decomp}.
	
	Following a similar argument, we can show that $$2\EE{C(Z)(\tau'(X)-\tau(X)) \p{\frac{\partial B_{t_\Delta}(Z) \varsigma(X)}{\partial t_\Delta}\bigg|_{t_\Delta=0}}}=0.$$
	
	Note that 
	\begin{align*}
		\frac{\partial C_{t_\Delta}(Z) \tau(X)}{\partial t_\Delta}\bigg|_{t_\Delta=0} &= - \frac{\partial A_{t_\Delta} (Z)}{\partial t_\Delta}\bigg|_{t_\Delta=0} \p{s(X) + \frac{\Delta(X)}{t(X)}} - A(Z)  \p{\frac{\Delta'(X)-\Delta(X)}{t(X)}}\tau(X) \\
		& \ \ \  - \frac{\partial B_{t_\Delta} (Z)}{\partial t_\Delta}\bigg|_{t_\Delta=0} \p{t(X) + \frac{\Delta(X)}{s(X)}} - B(Z)  \p{\frac{\Delta'(X)-\Delta(X)}{s(X)}}\tau(X).
	\end{align*}
	It's immediate from the previous arguments that  $$2\EE{C(Z)\p{\frac{\partial B_{t_\Delta}(Z) }{\partial t_\Delta}\bigg|_{t_\Delta=0}}\cond X}=0$$ and 
	$$2\EE{C(Z)\p{\frac{\partial A_{t_\Delta}(Z)}{\partial t_\Delta}\bigg|_{t_\Delta=0}}\cond X}=0,$$ and from Proposition \ref{prop:decomp}, again we use the fact that  and $\EE{C(Z)A(Z)\cond X}=0$ and \newline$\EE{C(Z)B(Z)\cond X}=0$, we then conclude that $$2\EE{C(Z)(\tau'(X)-\tau(X)) \p{\frac{\partial C_{t_\Delta}(Z) \tau(X)}{\partial t_\Delta}\bigg|_{t_\Delta=0}}}=0.$$ We can then conclude $$
	D_\Delta D_\tau L(\tau, \{m(x), \nu(x), \varsigma(x),  s(x), t(x), \Delta(x)\})[\tau' - \tau, \Delta'-\Delta] =0.$$
	
	An almost identical derivation can be taken to show that $$
	D_s D_\tau L(\tau, \{m(x), \nu(x), \varsigma(x),  s(x), t(x), \Delta(x)\})[\tau' - \tau, s'-s] =0.$$ We have now concluded showing Assumption 2 from \cite{foster2019orthogonal} holds.

  To show Assumption 3 holds, we have
	\begin{align*}
		&\frac{d^2}{dt_1t_2} \EE{(Y-m(X) - A(Z)\nu(X)-B(Z)\varsigma(X)- C(Z)(\bar{\tau}(X) + t_1 (\tau'-\tau) + t_2(\tau'-\tau)))^2}\\
		&= 2C(Z)^2(\tau'-\tau)^2
	\end{align*}
Given $s(X) + \frac{\Delta(X)}{t(X)} = \Omega(\eta)$ and $\abs{A(Z)} = \Omega(\eta)$, etc.,  we have $C(Z)=\Omega(\eta^2)$. Thus, $\lambda = \Omega(\eta^4)$ and $\kappa=0$ in Assumption 3. 

To show Assumption 4 holds, we can check that 
\begin{align*}
&	D_m^2D_\tau L(\tau, \{\bar{m}(x), \nu(x), \varsigma(x),  s(x), t(x), \Delta(x)\})[\tau' - \tau, m'-m, m'-m] =0.\\
&		D_\nu^2D_\tau L(\tau, \{m(x), \bar{\nu}(x), \varsigma(x),  s(x), t(x), \Delta(x)\})[\tau' - \tau, \nu'-\nu, \nu'-\nu] =0.\\
	&			D_\varsigma^2D_\tau L(\tau, \{m(x), \nu(x),\bar{ \varsigma}(x),  s(x), t(x), \Delta(x)\})[\tau' - \tau, \varsigma'-\varsigma, \varsigma'-\varsigma] =0.
\end{align*}
Finally, we can check that 
\begin{align*}
	&	D_\Delta^2D_\tau L(\tau, \{m(x), \nu(x), \varsigma(x),  s(x), t(x), \bar{\Delta}(x)\})[\tau' - \tau, \Delta'-\Delta, \Delta'-\Delta] \\
	&=O(M^2/\eta^4)\Norm{\tau'-\tau}_{L_2(P)}\Norm{\Delta'-\Delta}_{L_2(P)}
\end{align*}
and this similarly holds when we take derivative with respect to $s(X)$ or $t(X)$. We conclude that $\beta_2 = O(M^2/\eta^4)$ in Assumption 4. 
	
	The result then directly follows from Theorem 1 in \citet{foster2019orthogonal}.

\end{document}